\documentclass[12pt]{article}
\usepackage{amsfonts}
\usepackage{amsmath}
\usepackage{mathabx}
\usepackage{enumerate}
\usepackage[margin=1in]{geometry}
 \usepackage{graphicx} 
 \usepackage{algpseudocode}
 \usepackage{algorithm}
 \usepackage{color}
 \usepackage{wasysym}
  \usepackage[numbers]{natbib} 
  \usepackage{rotating}
\usepackage{amsthm}
\usepackage{fullpage}
\usepackage{hyperref}

\bibpunct{(}{)}{,}{a}{,}{,}

\newtheorem{proposition}{Proposition}
\newtheorem{example}{Example}
\newtheorem*{proposition*}{Proposition}

\DeclareMathOperator*{\CRM}{CRM}
\DeclareMathOperator*{\NRM}{NRM}
\DeclareMathOperator*{\PK}{PK}
\DeclareMathOperator*{\NGG}{NGG}
\def\PP{\mathbb{P}}

 \def\simind{\stackrel{\mbox{\scriptsize{ind}}}{\sim}}
\def\simiid{\stackrel{\mbox{\scriptsize{iid}}}{\sim}}

\title{\bf{A marginal sampler for $\sigma$-stable Poisson-Kingman mixture models}}

\date{\today}
\author{
 Mar\'ia Lomel\'i\\  Gatsby Computational Neuroscience Unit, UCL\\
   \and
 Stefano Favaro\\
  Department of Economics and Statistics, University of Torino\\ and Collegio Carlo Alberto\\
  \and
  Yee Whye Teh\\ Department of Statistics, University of Oxford\\}

\begin{document}
\def\spacingset#1{\renewcommand{\baselinestretch}%
{#1}\small\normalsize} \spacingset{1}

\graphicspath{ {./figures/} }
\maketitle

\begin{abstract}
\normalsize{We investigate the class of $\sigma$-stable Poisson-Kingman random probability measures (RPMs) in the context of Bayesian nonparametric mixture modeling. This is a large class of discrete RPMs which encompasses most of the popular discrete RPMs used in Bayesian nonparametrics, such as the Dirichlet process, Pitman-Yor process, the normalized inverse Gaussian process and the normalized generalized Gamma process. We show how certain sampling properties and marginal characterizations of $\sigma$-stable Poisson-Kingman RPMs can be usefully exploited for devising a Markov chain Monte Carlo (MCMC) algorithm for making inference in Bayesian nonparametric mixture modeling. Specifically, we introduce a novel and efficient  MCMC sampling scheme in an augmented space that has a fixed number of auxiliary variables per iteration. We apply our sampling scheme for a density estimation and clustering tasks with unidimensional and multidimensional datasets, and compare it against competing sampling schemes.
}

 \end{abstract}

\spacingset{1.45}
\section{Introduction}

Flexibly modeling the distribution of continuous data is an important concern in Bayesian nonparametrics and it requires the specification of a prior model for continuous distributions.  A fruitful and general approach for defining such a prior model was first suggested by \citet{Lo1984a} in terms of an infinite dimensional mixture model, which is nowadays the subject of a rich and active literature. Specifically, let $P$ be a discrete random probability measure (RPM) with distribution $\mathcal{P}$. Given a collection of continuous and possibly multivariate observations $X_{1},\ldots,X_{n}$, the infinite dimensional mixture model is defined hierarchically by means of a corresponding collection of latent random variables $Y_1,\ldots,Y_n$ from an exchangeable sequence directed by $P$, i.e., 
\begin{align}\label{eq:hierarchical_intro}
P& \quad\sim\quad \mathcal{P},\notag\\
Y_i|P &\quad\simiid\quad P\notag\\[4pt]
X_i|Y_i &\quad\simind\quad F(\cdot |Y_i)
\end{align}
where $F(\cdot|Y_i)$ is a continuous distribution parameterized by $Y_{i}$. The distribution $F(\cdot|Y_i)$ is referred to as the kernel, whereas $P$ is the mixing measure. The nonparametric hierarchical model \eqref{eq:hierarchical_intro} defines a mixture model with a countably infinite number of components. By the discreteness of $P$, each pair of the $Y_{i}$'s takes on the same value with positive probability, with this value identifying a mixture component. In such a way, the $Y_{i}$'s allocate the $X_{i}$'s to a random number of mixture components, thus providing a model for the unknown number of clusters within the data. The formulation of \eqref{eq:hierarchical_intro} presented in  \citet{Lo1984a} sets $P$ to be the Dirichlet process introduced by \citet {Fer1973a}, hence the name of Dirichlet process mixture model.

It is apparent that one can replace a Dirichlet process mixing measure with any other discrete RPM.  \citet{IshJam2001a} first replaced the Dirichlet process with stick-breaking RPMs. As a notable example they focussed on the two parameter Poisson-Dirichet process, also known as Pitman-Yor process, which is a discrete RPM introduced in \citet{PerPitYor1992a} and further investigated in \citet{PitYor1997a} and \citet{Jam2002a}. \citet{NiePruWal2004a} replaced the Dirichlet process with normalized random measures (NRMs) and \citet{LijMenPru2007a} focused on the normalized generalized Gamma process. See also \citet{Jam2002a} \citet{LijMenPru2005a}, \citet{JamLijPru2009a} and \citet{Jam2013a}. Both the Pitman-Yor process and the normalized generalized Gamma process are valid alternatives to the Dirichlet process: they preserve almost the same mathematical tractability but they also provide clustering properties that make use of all of the information contained in the sample. It is well known that the Dirichlet process allocates observations to the mixture model components with a probability depending solely on the number of times that the mixture's component occurs. In contrast, under the Pitman-Yor process and the normalized generalized Gamma process, the allocation probability depends heavily on the number of distinct mixture components. This more flexible allocation mechanism turns out to be a key feature for making inference under the mixture model \eqref{eq:hierarchical_intro}. See \citet{LijMenPru2005a} and \citet{LijMenPru2007a} for details.

Several Markov chain Monte Carlo (MCMC) methods have been proposed for posterior sampling from the Dirichlet process mixture model. On one hand, the marginal MCMC methods remove the infinite dimensionality of the problem by exploiting the tractable marginalization with respect to the Dirichlet process. See \citet{Esc1994a}, \citet{Mac1994a} and \citet{EscWes1995a} for early works, and \citet{Nea2000a} for an overview with some noteworthy developments such as the celebrated Algorithm 8. On the other hand, the conditional MCMC methods maintain the infinite dimensional part and find appropriate ways for sampling a sufficient but finite number of the atoms of the Dirichlet process. See \citet{IshJam2001a}, \citet{Wal2007a}, \citet{PapRob2008a}. Recently, marginal and conditional MCMC methods have been developed under more general classes of mixing measures, such as stick-breaking RPMs and NRMs, among others. See \citet{IshJam2001a}, \citet{GriWal2011a}, \citet{BarLijNie2013a}, \citet{FavTeh2013a} and \citet{FavLomTeh2014a} for details.

In this paper we introduce a marginal MCMC method for posterior sampling from \eqref{eq:hierarchical_intro} with $P$ belonging the class of $\sigma$-stable Poisson-Kingman RPMs introduced in \citet{Pit2003a}. We refer to such a model as a $\sigma$-stable Poisson-Kingman mixture model. A conditional MCMC method for $\sigma$-stable Poisson-Kingman mixture model has been recently introduced in \citet{favaroslice12}. The class of $\sigma$-stable Poisson-Kingman RPMs forms a large class of discrete RPMs which encompasses most of the popular discrete RPMs used in Bayesian nonparametrics, e.g., the Pitman-Yor and the normalized generalized Gamma processes. It also includes the Dirichlet process as a special limiting case. Our main contribution is to provide a general framework for doing posterior inference with all the members of this class of priors. Differently from  \citet{favaroslice12}, we exploit marginal properties of $\sigma$-stable Poisson-Kingman RPMs in order to remove the infinite dimensionality of the sampling problem. Since efficient algorithms often rely upon simplifying properties of the priors, just as inference algorithms for graphical models rely upon the conditional independencies encoded by the graph, in our experiments, we found that this improved the algorithmic performance. We applied our algorithm for a density estimation and clustering tasks with unidimensional and multidimensional datasets and compare it against competing sampling schemes.

 The paper is structured as follows. In Section 2, we recall the definition of $\sigma$-stable Poisson-Kingman RPM, as well as some of its marginal properties which are fundamental for devising our marginal MCMC method. In Section 3, we present the marginal MCMC method for posterior sampling $\sigma$-stable Poisson-Kingman mixture models. Section 4 contains unidimensional and multidimensional experiments and Section 5 concludes with a brief discussion.

\section{Preliminaries on $\sigma$-stable Poisson-Kingman RPMs}

We start by recalling the definition of completely random measures (CRMs), the reader is referred to \citet{Kin1967a} for a detailed account on CRMs. Let $\mathbb{X}$ be a complete and separable metric space endowed with the Borel $\sigma$-field $\mathcal{B}( \mathbb{X})$. A CRM $\mu$ is a random element taking values on the space of boundedly finite measures on $\mathbb{X}$ such that, for any $A_{1},\ldots,A_{n}$ in $\mathcal{B}( \mathbb{X} )$, with $A_{i}\cap A_{j}=\emptyset$ for $i\neq j$, the random variables $\mu(A_{1}),\ldots,\mu(A_{n})$ are mutually independent. \citet{Kin1967a} showed that $\mu$ is discrete almost surely so it can be represented in terms of nonnegative random masses $(u_k)_{k\geq1}$ at $\mathbb{X}$-valued random locations $(\phi_k)_{k\geq1}$, that is $\mu=\sum_{k\geq1}u_{k}\delta_{\phi_{k}}$. The distribution of $\mu$ is characterized in terms of the distribution of the random point set $(u_{k},\phi_{k})_{k\geq1}$ as a Poisson random measure on $\mathbb{R}^{+}\times\mathbb{X}$ with mean measure $\nu$, which is referred to as the L\'evy intensity measure. In this paper we focus on homogeneous CRMs, namely, CRMs such that $\nu(ds,dy)=\rho(ds)H_{0}(dy)$ for some L\'evy measure $\rho$ on $\mathbb{R}^{+}$ and some non-atomic base distribution $H_{0}$ on $\mathbb{X}$. Homogeneity implies independence between $(u_k)_{k\ge 1}$ and $(\phi_{k})_{k\ge 1}$, where $(\phi_{k})_{k\ge 1}$ are independent and identically distributed as $H_{0}$ while the law of $(u_k)_{k\ge 1}$ is governed by $\rho$ and denote by $\CRM(\rho,H_0)$ the distribution of a homogeneous CRM.

Homogeneous CRMs provide a fundamental tool for defining almost surely discrete random probability measures (RPMs) via the normalization approach. Specifically, let $\mu$ be a homogeneous CRM with L\'evy measure $\rho$ and base distribution $H_0$. Furthermore, let $T=\mu(\mathbb{X})=\sum_{k\geq1}u_{k}$ be the total mass of $\mu$. Both positiveness and finiteness of the random variable $T$ are ensured by the following conditions: $\int_{\mathbb{R}^{+}}\rho(ds)=+\infty$ and $\int_{\mathbb{R}^{+}}(1-\text{e}^{-s})\rho(ds)<+\infty$. Once these conditions are satisfied, one can define an almost surely discrete RPM as
\begin{align}\label{eq:rpm}
P =\frac{\mu}{T}=\sum_{k\geq1}{p_k\delta_{\phi_{k}}}
\end{align}
with $p_{k}=u_{k}/T$. Since $\mu$ is homogeneous, the law of the random probabilities $(p_k)_{k\ge 1}$ is governed by the L\'evy measure $\rho$, and the atoms $(\phi_k)_{k\geq1}$ are random variables independent of $(p_k)_{k\geq1}$ and independent and identically distributed according to $H_{0}$. The RPM displayed in \eqref{eq:rpm} is known from \citet{JamLijPru2009a} as the normalized random measure (NRM) with L\'evy measure $\rho$ and base distribution $H_0$. We refer to \citet{Jam2002a} and \citet{RegLijPru2003a} for a comprehensive account on homogeneous NRMs and denote by $\NRM(\rho,H_0)$ the distribution of $P$.

Since $P=\mu/T$ is almost surely discrete, there is positive probability of $Y_i=Y_j$ for each pair of indexes $i\neq j$.  This induces a random partition $\Pi$ on $\mathbb{N}$, where $i$ and $j$ are in the same block in $\Pi$ if and only if $Y_i=Y_j$.  \citet{Kin1978a} showed that $\Pi$ is exchangeable and its distribution, the so-called exchangeable partition probability function (EPPF), can be deduced from the law of the NRM. See \citet{Pit2006a} for a comprehensive account of EPPFs. A  second object induced by $(Y_i)_{i\ge 1}$ is a size-biased permutation of the atoms in $\mu$.  Specifically, order the blocks in $\Pi$ by increasing order of the least element in each block, and for each $k\in\mathbb{N}$ let $Z_k$ be the least element of the $k$th block.  $Z_k$ is the index among $(Y_i)_{i\ge 1}$ of the first appearance of the $k$th unique value in the sequence.  Let $V_k=\mu(\{Y_{Z_k}\})$ be the mass of the corresponding atom in $\mu$.  Then $(V_k)_{k\ge 1}$ is a size-biased permutation of the masses of atoms in $\mu$, with larger masses tending to appear earlier in the sequence.  It is easy to see that $\sum_{k\ge 1}V_k=T$, and that the sequence can be understood as a stick-breaking construction: starting with a stick of length $S_0=T$; break off the first piece of length $V_1$; the leftover length of stick is $S_1=S_0-V_1$; then the second piece with length $V_2$ is broken off, etc. 

Theorem 2.1 of \citet{PerPitYor1992a} states that the sequence of surplus masses  $(S_k)_{k\ge 0}$ forms a Markov chain and gives the corresponding initial distribution and transition kernels, see the supplementary material for details. Let us denote by $f_\rho(t)$ the density function of $T$. The EPPF of the random partition $\Pi$ can be derived from this theorem by enriching the generative process for the sequence $(Y_i)_{i\ge 1}$ as follows, where we simulate parts of the CRM as and when required.  
\begin{itemize}
\item[i)] Start with drawing the total mass from its distribution $\PP_{\rho,H_0}(T\in dt) = f_\rho(t)dt$.
\item[ii)] The first draw $Y_1$ from $\mu/T$ is a size-biased pick from the masses of $\mu$.  The actual value of $Y_1$ is simply $Y^*_1\sim H_0$, while the mass of the corresponding atom in $\mu$ is $V_1$, with conditional distribution given by
$$\PP_{\rho,H_0}(V_1\in dv_1|T\in dt) = \frac{v_1}{t}\rho(dv_1) \frac{f_\rho(t-v_1)}{f_\rho(t)}.$$
The leftover mass is $S_1=T-V_1$.
\item[iii)] For subsequent draws $i\geq2$:
\begin{itemize}
\item Let $K$ be the current number of distinct values among $Y_1,\ldots,Y_{i-1}$, and $Y^*_1,\ldots,Y^*_K$ the unique values, i.e., atoms in $\mu$.  The masses of these first $K$ atoms are denoted $V_1,\ldots,V_K$ and the leftover mass is $S_K=T-\sum_{k=1}^K V_k$.
\item For each $k\le K$, with probability $V_k/T$, we set $Y_i=Y^*_k$.
\item With probability $S_K/T$, $Y_i$ takes on the value of an atom in $\mu$ besides the first $K$ atoms.  The actual value  $Y^*_{K+1}$ is drawn from $H_0$, while its mass is
drawn from
$$\PP_{\rho,H_0}(V_{K+1}\in dv_{K+1}|S_K\in ds_K) 
= \frac{v_{K+1}}{s_K}\rho(dv_{K+1}) \frac{f_\rho(s_K-v_{K+1})}{f_\rho(s_K)}.$$
The leftover mass is again $S_{K+1}=S_K-V_{K+1}$.
\end{itemize}
\end{itemize}
By multiplying the above infinitesimal probabilities one obtains the joint distribution of the random elements $T$, $\Pi$,  $(V_{i})_{i\geq1}$ and $(Y^{\ast}_{i})_{i\geq1}$. Such a joint distribution was first obtained in \citet{Pit2003a} and it is recalled in the next proposition, see also \citet{Pit2006a} for details.

\begin{proposition}\label{crmgenprob}
Let $\Pi_n$ be the exchangeable random partition of $[n]:=\{1,\ldots,n\}$ induced by a sample $(Y_i)_{i\in [n]}$ from $P\sim \NRM(\rho,H_0)$.   Let  $(Y_j^*)_{j\in[K]}$ be the $K$ distinct values in $(Y_i)_{i\in [n]}$ with masses $(V_j)_{j\in[K]}$.  Then
\begin{align}\label{dist_prop}
&\PP_{\rho,H_0}(T\in dt, \Pi_n=(c_k)_{k\in[K]}, Y_k^*\in dy_k^*, V_k\in dv_k\text{ for }k\in[K])\\
&\notag\quad= t^{-n}f_\rho(t-\textstyle\sum_{k=1}^{K} v_k) dt
\displaystyle \prod_{k=1}^{K} v_k^{|c_k|} \rho(dv_k)H_0(dy_k^*),
\end{align}
where $(c_k)_{k\in[K]}$ denotes a particular partition of $[n]$ with $K$ blocks, $c_1,\ldots,c_K$, ordered by increasing least element  and $|c_k|$ is the cardinality of block $c_k$. The distribution \eqref{dist_prop} is invariant to the size-biased order.
\end{proposition}

\subsection{$\sigma$-stable Poisson-Kingman RPMs}

Poisson-Kingman RPMs have been introduced in \citet{Pit2003a} as a generalization of homogeneous NRMs. Let $\mu\sim\CRM(\rho,H_0)$ and let $T=\mu(\mathbb{X})$ be finite, positive almost surely, and absolutely continuous with respect to Lebesgue measure. For any $t\in\mathbb{R}^+$, let us consider the conditional distribution of $\mu/t$ given that the total mass $T\in dt$. This distribution is denoted by $\PK(\rho,\delta_t,H_0)$, it is the distribution of a RPM and $\delta_t$ denotes the Dirac delta function. Poisson-Kingman RPMs form a class of RPMs whose distributions are obtained by mixing $\PK(\rho,\delta_t,H_0)$, over $t$, with respect to some distribution $\gamma$ on the positive real line. Specifically, a Poisson-Kingman RPM has the hierarchical representation
\begin{align}
T &\sim \gamma \nonumber \label{PKP}\\
P|T=t &\sim \PK(\rho,\delta_t,H_0).
\end{align}
The RPM $P$ is referred to as the Poisson-Kingman RPM with L\'evy measure $\rho$, base distribution $H_0$ and mixing distribution $\gamma$. Throughout the paper we denote by $\PK(\rho,\gamma,H_0)$ the distribution of $P$. If $\gamma(dt) = f_\rho(t)dt$ then the distribution $\PK(\rho,f_\rho,H_0)$ coincides with $\NRM(\rho,H_0)$.  Since $\mu$ is homogeneous, the atoms $(\phi_k)_{k\ge 1}$ of $P$ are independent of their masses $(p_k)_{k\ge 1}$ and form a sequence of independent random variables identically distributed according to $H_{0}$. Finally, the masses of $P$ have distribution governed by the L\'evy measure $\rho$ and the distribution $\gamma$.

In this paper we focus on the class of $\sigma$-stable Poisson-Kingman RPMs. This is a noteworthy subclass of Poisson-Kingman RPMs which encompasses most of the popular discrete RPMs used in Bayesian nonparametrics, e.g., the Pitman-Yor process and the normalized generalized Gamma process. For any $\sigma\in(0,1)$, $f_{\sigma}(t) = \frac{1}{\pi}\sum_{j=0}^{\infty}{\frac{(-1)^{j+1}}{j!}}\textrm{sin}(\pi \sigma j)\frac{\Gamma(\sigma j +1)}{t^{\sigma j +1}}$ is the density function of a positive $\sigma$-stable random variable, let us denote it by $f_{\sigma}$. A $\sigma$-stable Poisson-Kingman RPMs is a Poisson-Kingman RPM with L\'evy measure
\begin{equation}\label{levymeas}
\rho(dx)=\rho_{\sigma}(dx):=\frac{\sigma}{\Gamma(1-\sigma)}x^{-\sigma-1}dx, 
\end{equation}
base distribution $H_0$ and mixing distribution $\gamma(dt)=h(t)f_{\sigma}(dt)/\int_{0}^{+\infty}h(t)f_{\sigma}(t)dt$, for any nonnegative measurable function $h$. Accordingly, $\sigma$-stable Poisson-Kingman RPMs form a class of discrete RPMs indexed by the parameter $(\sigma,h)$.  The Dirichlet process is a limiting $\sigma$-stable Poisson-Kingman RPM if $\sigma\rightarrow0$, for some choices of $h$. Throughout the paper we denote by $\PK(\rho_{\sigma},h,H_0)$ the distribution of a $\sigma$-stable Poisson-Kingman RPM with parameter $(\sigma,h)$.

\begin{figure}[h!]
\centering  
\includegraphics[scale=0.7]{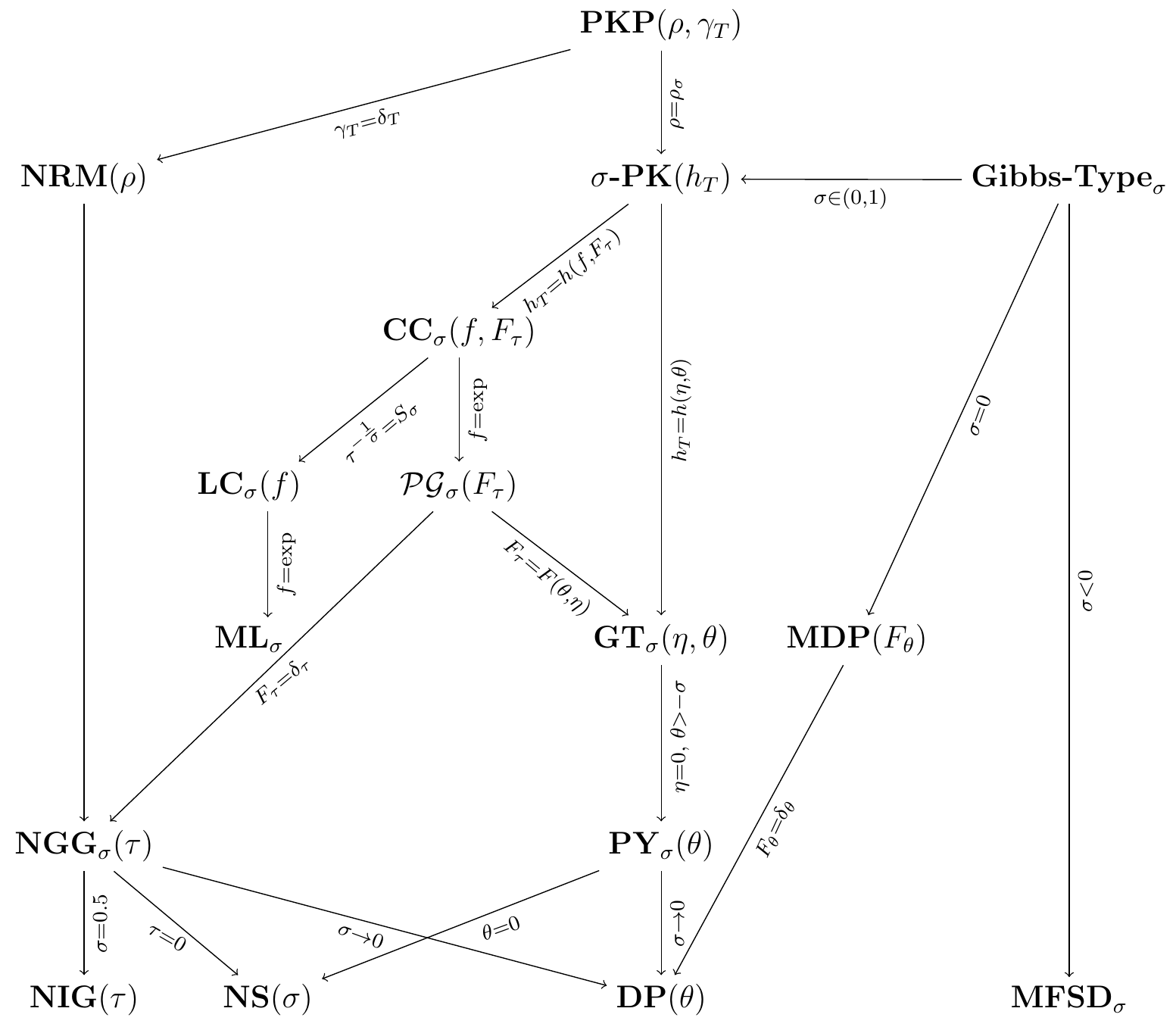}\caption{\normalsize{Mixture of DPs (MDP), Mixture of finite symmetric Dirichlet (MFSD).} \label{PKpriorsmap}}
\end{figure}

Examples of $\sigma$-stable Poisson-Kingman RPMs are obtained by specifying the tilting function $h$. The \textit{normalized $\sigma$-stable process} (NS) in \citet{Kin1975a} corresponds to $h(t)=1$. The \textit{normalized generalized gamma process} (NGG) in \citet{Jam2002a} and \citet{Pit2003a} corresponds to $ h(t)=\exp\{\tau-\tau^{1/\sigma} t\}$, for any $\tau >0$. See also \citet{LijMenPru2005a}, \citet{LijMenPru2007a}, \citet{lijoi08}, \citet{JamLijPru2009a} and \citet{Jam2013a}. The \textit{Pitman-Yor process} (PY) in \citet{PerPitYor1992a} corresponds to 
$h(t)=\frac{\Gamma(\theta+1)}{\Gamma(\theta/\sigma+1)}t^{-\theta}$ with $\theta\geq -\sigma$,
see \citet{PitYor1997a}.  The \textit{Gamma-tilted} process (GT) corresponds to $h(t)=t^{-\theta}\exp\{-\eta t \}$, for any $\eta>0$ or $\eta=0$ and $\theta>-\sigma$.  The \textit{Poisson-Gamma class} ($\mathcal{PG}$) in \citet{Jam2013a} corresponds to $h(t)=\int_{\mathbb{R}^{+}}\exp\{\tau-\tau^{1/\sigma} t\}F(d\tau)$, for any distribution function $F$ over the positive real line. See also \citet{PitYor1997a} and \citet{Jam2002a}.  Let $\mathcal{T}$ a positive random variable, the \textit{composition of classes} (CC) in \citet{james08} corresponds to 
$h(t)=\frac{\mathbb{E}[f(t\mathcal{T}^{1/\sigma})]}{\int_{0}^{+\infty}{\mathbb{E}[f(t\mathcal{T}^{1/\sigma})]f_{\sigma}(t)\textrm{d}t}}$,
where $f$ is any positive function, see \citet{Jam2002a} for details. Let $S_\sigma$ a positive $\sigma$-stable random variable, the \textit{Lamperti class} (LT) in \citet{james08} corresponds to the choice 
 \begin{equation}\label{lamp}
h(t)= L_{\sigma}\mathbb{E}\left(g( S_{\sigma}t^{-1})\right),
\end{equation}
where%
$\frac{1}{L_{\sigma}}= \frac{\sin(\pi\sigma)}{\pi} \int_{0}^{+\infty}{\frac{f(y)y^{\sigma-1}}{y^{2\sigma}+2y^{\sigma}\cos(\pi \sigma)+1}\textrm{d}y}$
and $g$ is any positive function such that \eqref{lamp} is well-defined, see \citet{Jam2002a} for details. The Mittag-Leffler class (ML) in \citet{james08} corresponds to the choice of $g(x)=\exp\{- x^{\sigma}\}$ in the tilting function \eqref{lamp}. Figure \ref{PKpriorsmap} shows the relationships among these examples of $\sigma$-stable Poisson-Kingman RPMs. 

The distribution of the exchangeable random partition induced by a sample from a $\sigma$-stable Poisson-Kingman RPMs is obtained by a direct application of Proposition \ref{crmgenprob}. See the supplementary material for details about the EPPF under $\sigma$-stable Poisson-Kingman RPMs. The next proposition specializes Proposition \ref{crmgenprob} to the context of $\sigma$-stable Poisson-Kingman RPMs.

\begin{proposition}\label{prop2}
Let $\Pi_n$ be the exchangeable random partition of $[n]$ induced by a sample $(Y_i)_{i\in [n]}$ from $P\sim\PK(\rho_\sigma,h,H_0)$. Then,
\begin{equation}\label{gibbs_eppf}
\PP_{\rho_\sigma,h,H_0}(\Pi_n=(c_k)_{k\in[K]})= V_{n,K}\prod_{k=1}^{K} W_{|c_k|}
\end{equation}
where we set $V_{n,K}= \int_{\mathbb{R}^+}\int_{0}^{t} t^{-n}(t-s)^{n-1-k\sigma}h(t)f_\sigma(s)dt ds$ and $W_m = \Gamma(m-\sigma)/\Gamma(1-\sigma)= [1-\sigma]_{m-1}:=\prod_{i=0}^{m-2}(1-\sigma+i)$.
\end{proposition}

Proposition \ref{prop2} provides one of the main tools for deriving the marginal MCMC sampler in Section 3.
We refer to \citet{GnePit2006a} and \citet{Pit2006a} for a comprehensive study of exchangeable random partitions with distribution of the form \eqref{gibbs_eppf}. These random exchangeable partitions are typically referred to as Gibbs-type with parameter $\sigma\in(0,1)$.

\section{Marginal samplers for $\sigma$-stable Poisson-Kingman mixture models}\label{sampler}

In this section we develop a marginal sampler that can be effectively applied to all members of the $\sigma$-stable Poisson-Kingman process family.  Our sampler does not require any numerical integrations, nor evaluations of special functions, e.g.\ the density $f_\sigma$ of the positive $\sigma$-stable distribution as in \citet{WueMaeRme2013a}.  It applies to non-conjugate hierarchical mixture models based on $\sigma$-stable Poisson-Kingman RPMs, by extending the Reuse data augmentation scheme of \citet{FavTeh2013a}.

\subsection{Effective representation with data augmentation}

If $\gamma(dt)\propto h(t) f_\sigma(t)$, the joint distribution over the induced partition $\Pi_n$, total mass $T$ and surplus mass $S$ is given by:
\begin{align}
&\PP_{\rho_\sigma,\gamma,H_0}(T\in dt, S\in ds, \Pi_n=(c_k)_{k\in[K]}) \nonumber \\
=& t^{-n}(t-s)^{n-1-K\sigma} f_\sigma(s) h(t) dt
\displaystyle \frac{\sigma^{K}}{\Gamma(n-K\sigma)}\prod_{k=1}^{K} \frac{\Gamma(|c_k|-\sigma)}{\Gamma(1-\sigma)} \mathbb{I}_{(0,t)}(s)\mathbb{I}_{(0,\infty)}(t).\nonumber
\end{align}
Except for two difficulties, this joint distribution easily allows us to construct marginal samplers.  The first difficulty is that it is necessary to compute $f_\sigma$ if working with an MCMC scheme using the above representation.  Current software packages compute these using numerical integration techniques, which can be unnecessarily expensive. The following is an integral representation \citep{Kan1975a,zolotarev66}, with a view to introducing an auxiliary variable into our system thus removing the need to evaluate the integral numerically. %
Let $\sigma\in(0,1)$.  Then,
\begin{align}
f_\sigma(t)&= \frac{1}{\pi}\frac{\sigma}{1-\sigma}\left(\frac{1}{t}\right)^{\frac{1}{1-\sigma}}\int_0^{\pi}{A(z)\exp{\left(-\left(\frac{1}{t}\right)^{\frac{\sigma}{1-\sigma}}A(z)\right)}d z}
\label{zolotarev}
\end{align}
where Zolotarev's function is
\[
A(z)= \left[\frac{\sin(\sigma z)}{\sin(z)}\right]^{\frac{1}{1-\sigma}}\left[\frac{\sin((1-\sigma)z)}{\sin(\sigma z)}\right],\hspace{2cm}z\in(0,\pi).
\]

Zolotarev's representation has been used by  \citet{devroye} to construct a random number generator for polynomially and exponentially tilted $\sigma$-stable random variables and in a rejection sampling scheme by \citet{favaroslice12}. 
Our proposal here is to introduce an auxiliary variable $Z$ using a data augmentation scheme (\citet{tanner87}), with conditional distribution given $T\in dt$ described by the integrand in~\eqref{zolotarev}.

The second difficulty is that the variables $T$ and $S$ are dependent and that computations with small values of $T$ and $S$ might not be numerically stable.  To address these problems, we propose the following reparameterization: $W = \frac{\sigma}{1-\sigma}\log T$, and $R=S/T$ where $W\in \mathbb{R}$  and $R\in (0,1)$.  This gives our final representation:
\begin{align}
&\PP_{\rho_\sigma,\gamma,H_0}(W\in dw, R\in dr, Z\in dz, \Pi_n=(c_k)_{k\in[K]})  \nonumber \\
=& 
\quad \frac{1}{\pi} e^{-w(1+(1-\sigma)K)} (1-r)^{n-1-K\sigma} r^{-\frac{1}{1-\sigma}} h(e^{\frac{1-\sigma}{\sigma}w}) A(z) e^{-e^{-w}r^{-\frac{\sigma}{1-\sigma}} A(z)}\frac{\sigma^K}{\Gamma(n-\sigma K)}\prod_{k=1}^K [1-\sigma]_{|c_k|-1}.
\nonumber 
\end{align}

\subsection{$\sigma$-stable Poisson-Kingman mixture models}

To make the derivation of our sampler explicit,  we will consider a $\sigma$-stable Poisson-Kingman RPM as the random mixing distribution in a Bayesian nonparametric mixture model:
\begin{align}
T&\sim \gamma \nonumber \\
P|T &\sim \PK(\rho_{\sigma}, \delta_T, H_0)\nonumber\\
Y_i \mid P &\stackrel{_{\textrm{iid}}} {\sim}P\nonumber\\
X_i \mid Y_i &\stackrel{\textrm{ind}}{\sim} F(\cdot \mid Y_i)\nonumber
\end{align}
where $F(\cdot \mid Y)$ is the observation's distribution, and our dataset consists of  $n$ observations $(x_i)_{i\in[n]}$ of the corresponding variables $(X_i)_{i\in[n]}$.  We will assume that $F(\cdot \mid Y)$ is smooth.

In the following we will derive two marginal samplers for our nonparametric mixture models.  As opposed to conditional samplers, which maintain explicit representations of the random probability measure $P$,  marginal samplers  marginalize out $P$, retaining only the induced partition $\Pi_n$ of the dataset. In our case, including as well the auxiliary variables $W$, $R$, $Z$ in the final representation presented above.  Denoting the unique values (component parameters) by $(Y_k^*)_{k\in [K]}$, the joint distribution over all variables is given by:
\begin{align}
&\PP(W\in dw, R\in dr, Z\in dz, \Pi_n=(c_k)_{k\in [K]},Y^*_k\in dy^*_k\text{ for } k\in[K], X_i\in dx_i\text{ for } i\in[n])   \nonumber \\
=& 
\frac{1}{\pi} e^{-w(1+(1-\sigma)K)} (1-r)^{n-1-K\sigma} r^{-\frac{1}{1-\sigma}} h(e^{\frac{1-\sigma}{\sigma}w}) A(z) e^{-r^{-\frac{\sigma}{1-\sigma}}e^{-w} A(z)}dw\,dr\,dz \nonumber \\
&\times\frac{\sigma^K}{\Gamma(n-\sigma K)}\prod_{k=1}^K [1-\sigma]_{n_k-1} H_0(dy^*_k) \prod_{i\in c_k} F(dx_i|y^*_k).
\label{joint} 
\end{align}

The system of predictive distributions governing the distribution over partitions given the other variables can be read from the joint distribution~\eqref{joint}.  Specifically, the conditional distribution of a new variable $Y_{n+1}$ is:
\begin{align}
\quad&\PP(Y_{n+1}\in dy_{n+1}\,|\,\Pi_n=(c_k)_{k\in[K]},Y^*_k\in dy^*_k\text{ for } k\in[K],W\in dw,R\in dr,Z\in dz) 
\nonumber \\
\quad \propto\,& \sigma e^{(\sigma-1)w}(1-r)^{-\sigma}\frac{\Gamma(n+1-\sigma K)}{\Gamma(n+1-\sigma(K+1))} H_0(dy_{n+1}) + \sum_{k=1}^K (|c_k|-\sigma) \delta_{y^*_k}(dy_{n+1}).\nonumber
\end{align}
The conditional probability of the next observation joining an existing cluster $c_k$ is proportional to $|c_k|-\sigma$, which is the same for all exchangeable random probability measures based on the $\sigma$-stable CRM.  The conditional probability of joining a new cluster is more complex and dependent upon the auxiliary variables.
Such system of predictive distributions were first studied by \citet{blackwell73} for the chinese restaurant process, see also \citet{Ald1985a} for details and \citet{ewens72} for an early account in population genetics.

\subsection{Sampler updates}

In this section we will first describe Gibbs updates to the partition $\Pi_n$, conditioned on the auxiliary variables $W, R, Z$ before describing updates to the auxiliary variables.  We describe a non-conjugate case where the component parameters $(Y_k^*)_{k\in[K]}$ cannot be marginalized out and we derive an extension of \citet{FavTeh2013a}. In the case where the base distribution $H_0$ is conjugate to the observation's distribution $F(\cdot)$, the component parameters can be marginalized out as well, which leads to an extension to Algorithm 3 of \cite{Nea2000a}.

\subsubsection{Non-Conjugate Marginalized Sampler}

In general, the base distribution $H_0$ might not be conjugate to the observation's distribution $F$, and the cluster parameters cannot be marginalized out tractably.  In this case the state space of our Markov chain consists of $(c_k)_{k\in[K]}$, $W$, $R$, $Z$, as well as the cluster parameters $(Y^*_k)_{k\in[K]}$. The Gibbs updates to the partition involve updating the cluster assignment of one observation, say the $i$th one, at a time. We can adapt the Reuse algorithm of \citet{FavTeh2013a} to update our partitions. 

In this algorithm a fixed number $M>0$ of potential new clusters are maintained along with  those in the current partition $(c_k)_{k\in[K]}$. The parameters for each of these potential new clusters are denoted by $(Y^\text{new}_\ell)_{\ell\in[M]}$.  When updating the cluster assignment of the $i$th observation, we consider the potential new clusters as well as those in $(c_k^{\neg i})_{k\in[K^{\neg i}]}$.  If one of the potential new clusters is chosen, it is moved into the partition, and in its place a new cluster is generated by drawing a new parameter from $H_0$.  For the converse, when a cluster in the partition is emptied, it is moved into the list of potential new clusters, displacing a randomly chosen one.  Once every iteration through the dataset, the parameters of the potential new clusters are refreshed by  iid draws from $H_0$, see the pseudocode in Algorithm \ref{alg2} for details.  The conditional probability of the cluster assignment of the $i$th observation is:
\begin{align}%
\PP(i\text{ joins cluster $c_k^{\neg i}$ }|\text{ rest }) 
\propto& 
(|c_k^{\neg i}|-\sigma) F(X_i\in dx_i |y^*_k)  \nonumber\\
\PP(i\text{ joins new cluster $\ell$ }|\text{ rest }) 
\propto &
\frac{1}{M}\sigma e^{(\sigma-1)w}(1-r)^{-\sigma}
\frac{\Gamma(n-\sigma K^{\neg i})}{\Gamma(n-\sigma(K^{\neg i}+1))} 
F(X_i\in dx_i |y^\text{new}_\ell) \nonumber.
\end{align}
If $H_0$ is conjugate, we can replace the likelihood term in the cluster assignment rule by the conditional density of $x$ under cluster $c$, denoted by $F(X\in\textrm{d}x|\mathbf{x}_c)$, given the observations $\mathbf{X}_c=(X_j)_{j\in c}$ currently assigned to the cluster:
\begin{align}
F(X\in\textrm{d}x |\mathbf{x}_c) = \frac{\int H_0(dy) F(X\in \textrm{d}x|y)\prod_{j\in c} F(X	\in \textrm{d}x_j|y)}{\int H_0(dy) \prod_{j\in c} F(X\in \textrm{d}x_j|y)}.\nonumber
\end{align}

\subsubsection{Updates to Auxiliary Variables}

Updates to the auxiliary variables $W$, $R$ and $Z$ are straightforward.  Their conditional densities can be read off the joint density~\eqref{joint}:
\begin{align}
\PP(W\in dw\,|\text{ rest }) &\propto e^{-w(1+(1-\sigma)K)}h(e^{\frac{1-\sigma}{\sigma} w})e^{-r^{-\frac{\sigma}{1-\sigma}}e^{-w} A(z)} dw, \hspace{5mm}w\in \mathbb{R} \nonumber \\
\PP(R\in dr\,|\text{ rest }) &\propto (1-r)^{n-1-K\sigma} r^{-\frac{1}{1-\sigma}} e^{-r^{-\frac{\sigma}{1-\sigma}} e^{-w} A(z)} dr, \hspace{5mm} r\in (0,1) \nonumber\\
\PP(Z\in dz\,|\text{ rest }) &\propto A(z) e^{-r^{-\frac{\sigma}{1-\sigma}} e^{-w} A(z)} dz,\hspace{5mm} z\in (0,\pi)\nonumber.
\end{align}

\spacingset{1.45}

Although these are not in standard form, their states can be updated easily using generic MCMC methods.  We used slice sampling by \citet{Nea2003a} in our implementation, see the supplementary material for details.
If we have a prior on the index parameter $\sigma$, it can be updated as well.  Due to the heavy tailed nature of the conditional distribution, we recommend transforming $\sigma$ to $\log\frac{\sigma}{1-\sigma}$.

\spacingset{1}
\normalsize{
\begin{algorithm}\caption{ReUse$(\Pi_n,M,\{X_i\}_{i\in[n]},\{Y^*_c\}_{c\in\Pi_n},H_0)$}
 \label{alg2}
\begin{algorithmic}
\State Draw $\{Y_j^e\}_{j=1}^M \stackrel{\textrm{i.i.d.}}{\sim} H_0$
\For{$i= 1\to n$}
\State Let $c\in \Pi_n$ be such that $i\in c$
\State $c\leftarrow c\setminus\{i\}$
\If {$c=\emptyset$}
\State $k\sim\textrm{UniformDiscrete}(\frac{1}{M})$
\State $Y_k^e \leftarrow Y^*_c$
\State $\Pi_n\leftarrow \Pi_n\setminus \{c\}$
\EndIf
\State Set  $c'\textrm{ according to Pr}[i\textrm{ joins cluster }c'\mid \{X_i\}_{i\in c},Y^*_c, \textrm{rest}]$ 
\If{$c'\in[M]$ }
\State $\Pi_n\leftarrow \Pi_n\cup \{\{i\}\}$
\State $Y^*_{\{i\}}\leftarrow Y_{c'}^e$ 
\State $Y_{c'}^e\sim H_0$ 
\Else
\State $c' \leftarrow c' \cup \{i\}$ 
\EndIf
\EndFor
\end{algorithmic}
\end{algorithm}

\spacingset{1.45}

\subsection{	Differences with \cite{favaroslice12} conditional sampler}

If we start with Proposition \ref{crmgenprob}, we can do the following 2 changes of variables: $P_j=\frac{V_j}{T-\sum_{\ell<j}{V_{\ell}}}$ and $U_j=\frac{P_j}{1-\sum_{\ell<j}{P_{\ell}}}$. Then, we obtain the corresponding joint in terms of $N$ $(0,1)$-valued stick-breaking weights $\left\{U_j\right\}_{j=1}^N$ which corresponds to equation (19) of \cite{favaroslice12}. The truncation level $N$ needs to be randomised to have an exact MCMC scheme. The authors propose to do so with \cite{kalliwalker11}'s efficient slice sampler. The number of auxiliary variables, $N+2$, is a random quantity  as opposed to keeping it fixed when using our marginal sampler. This could potentially lead to slower running times and larger memory requirements to store these quantities when the number of data points is large. Furthermore, in our implementation of this sampler, we found that some of this auxiliary variables are highly correlated which leads to slow mixing of the chain. A quantitative comparison is presented in Table \ref{essgalaxymarg} in terms of running times and effective sample sizes (ESS).

\begin{algorithm}
\caption{MarginalSamplerNonConj($h_T,\sigma,M,H_0$)}
 \label{alg1} 
\begin{algorithmic}%
\For{$t = 2 \to iter$}
\State  Update $z^{(t)}$: Slice sample $\tilde{\mathbb{P}}\left( Z\in \textrm{d}z\mid \textrm{rest}\right)$ 
\State Update $p^{(t)}$ : Slice sample $\tilde{\mathbb{P}}\left( P\in \textrm{d}p \mid  \textrm{rest}\right)$
\State Update $w^{(t)}$: Slice sample $\tilde{\mathbb{P}}\left(W\in \textrm{d}w\mid  \textrm{rest}\right)$
\State Update \small{$\pi^{(t)}, \left\{x_c^*\right\}^{(t)} _{c\in \pi}$: \textbf{ReUse}$(\Pi_n,M,\{Y_i\}_{i\in[n]},\{X^*_c\}_{c\in\Pi_n},H_0\mid $rest)}
\EndFor 
\end{algorithmic}
\end{algorithm}}

 \section{Numerical illustrations}

\spacingset{1.45}
 
 In this section, we illustrate the algorithm on unidimensional and multidimensional data sets. We applied our MCMC sampler for density estimation of a $\sigma$-stable $\textrm{PK}\left(\sigma, H_0,h_T(t)\right)$ mixture model, for various choices of $h(t)$. In those experiments where we used a conjugate prior for the mixture component's parameter we sampled the parameters rather than integrating them out. We kept the hyperparameters of each $h$-tilting function fixed.

 \subsection{Unidimensional experiment}

  The dataset from  \citet{galaxy} consists of measurements of velocities in km/sec of 82 galaxies from a survey of the Corona Borealis region. We chose the base distribution $H_0$ and the corresponding likelihood $F$ for the $k$th cluster:
\begin{align}
H_0(\textrm{d}\mu_k, \text{d}\tau_k) =& F\left(\textrm{d}\mu_k\mid\mu_0,\tau_0\tau_k\right)F(\text{d}\tau_k\mid \alpha_0,\beta_0)\nonumber\\
F(\textrm{d}x_1,\hdots, \textrm{d}x_{n_k}\mid \mu_k,\tau_k) =& \prod_{i=1}^{n_k}\mathcal{N}\left(x_i\mid \mu_k, \tau_k^{-1}\right)\nonumber
\end{align}

\spacingset{1.45}{
where $X_1,\hdots,X_{n_k}$ are the observations currently assigned to cluster $k$. $\mathcal{N}$ denotes a Normal distribution with given mean $\mu_k$ and variance $\tau_k^{-1}$. In the first sampler (\textit{Marg-Conj I}), we used $H_0(\textrm{d}\mu_k, \text{d}\tau_k)=\mathcal{N}\left(\textrm{d}\mu_k\mid\mu_0,\tau_0^{-1}\right)\delta{\left\{\tau_k=\tau_1\right\}}$ with a common precision parameter among all clusters and set it to be $\frac{1}{4}$ of the range of the data. In the second sampler (\textit{Marg-Conj II}), we used $H_0(\textrm{d}\mu_k, \text{d}\tau_k) =\mathcal{N}\left(\textrm{d}\mu_k\mid\mu_0,\tau_0^{-1}\tau_k^{-1}\right)\text{Gamma}(\tau_k\mid \alpha_0,\beta_0)$. In the third sampler (\textit{Marg-NonConj}), we used a non conjugate distribution for the mean per cluster , $\mu = \log \varphi$ where $\varphi \sim \text{Gamma}(\varphi \mid a_0,b_0)$ and $\tau_k\sim\text{Gamma}(\tau_k\mid \alpha_0,\beta_0)$.

In  Table~\ref{cvsnc}, we reported a modest increase in the running times if we use a non-conjugate prior (\textit{Marg-NonConj}) for the mean versus a conjugate prior (\textit{Marg-Conj II}) . In Table~\ref{essgalaxymarg}, the algorithm's sensibility to the number of new components $M$ was tested and compared against the conditional sampler of \cite{favaroslice12}.} As we increase the marginal sampler's number of new components per iteration the ESS increases. Intuitively, the computation time increases but also leads to a potentially better mixing of the algorithm. In contrast, we found that the conditional sampler was not performing too well due to high correlations between the auxiliary variables. Finally, in Table~\ref{assess} we present that different values for $\sigma$ can be effectively chosen without modifying the algorithm as opposed to \cite{favaroslice12}, which is only available for $\sigma=0.5$.

\begin{table}
     \centering
      \spacingset{1}
  \scalebox{0.8}{
 \begin{tabular}{|ccccc|}
 \hline
\textbf{Algorithm}&$\mathbf{\sigma}$ &M&\textbf{Running time}&\textbf{ESS($\pm \textrm{std}$)}\\
 \hline
 Pitma-Yor process ($ \theta = 10$)&&&&\\
 \hline
Marginal-Conj II&0.5&4&23770.3(2098.22)&4857.644(447.583)\\
Marginal-NonConj&0.5&4&46352.4(252.27)& 5663.696(89.264)\\
\hline
Normalized Generalized Gamma process ($ \tau = 1$)&&&&\\
 \hline
Marginal-Conj II&0.5&4&22434.1(78.191)&3400.855(110.420)\\
Marginal-NonConj&0.5&4&28933.5(133.97)&5361.945(88.521)\\
\hline
 \end{tabular}}\caption{Running times in seconds and number of cluster's ESS averaged over 5 chains. Unidimensional dataset,  30,000 iterations, 10,000 burn in.\label{cvsnc}}
\end{table} 
 \begin{table}
   \centering
   \spacingset{1}
  \scalebox{0.8}{
 \begin{tabular}{|ccccc|}
 \hline
  \textbf{Algorithm}&$\mathbf{\sigma}$ &M &\textbf{Running time}&\textbf{ESS($\pm \textrm{std}$)}\\
 \hline
 Pitman-Yor process ($\theta = 50$)&&&&\\
 \hline
Marginal-Conj I&0.5&$2$&1.1124e+04&4121.94(821.562)\\
Marginal-Conj I&0.5&$6$&1.1216e+04&11215.55(596.249)\\ %
Marginal-Conj I&0.5&$10$&1.1355e+04&12469.87(548.981)\\
Marginal-Conj I&0.5&$15$ &1.1385e+04&13087.92(504.595)\\
Marginal-Conj I&0.5&$20$&1.1415e+04&12792.78(391.123)\\
 Conditional-Conj I&0.5&-&1.5659e+04&707.82 (95.754)\\
 \hline
Normalized Generalized Gamma process ($\tau = 50$)&&&&\\
 \hline
Marginal-Conj I&0.5&$2$&1.1617e+04&4601.63(574.339)\\
Marginal-Conj I&0.5&$6$&1.1650e+04&10296.85(425.333)\\
Marginal-Conj I&0.5&$10$&1.1692e+04&11415.41(377.418)\\
Marginal-Conj I&0.5&$15$&1.1795e+04&11473.44(374.031)\\
Marginal-Conj I&0.5&$20$&1.1875e+04&11461.08(506.744)\\
 Conditional-Conj I&0.5&-&1.5014e+04&848.73 (135.138)\\
\hline
 \end{tabular}}
 \caption{\normalsize{Running times in seconds and number of cluster's ESS averaged over 10 chains.  Unidimensional dataset, 50,000 iterations per chain, 20,000 burn in.}\label{essgalaxymarg}}
  \end{table}
  \begin{table}
     \centering
      \spacingset{1}
  \scalebox{0.8}{
 \begin{tabular}{|ccccc|}
 \hline
\textbf{Algorithm}&$\mathbf{\sigma}$ &$M$&\textbf{Running time}&\textbf{ESS($\pm \textrm{std}$)}\\
 \hline
 Pitma-Yor process ($ \theta = 10$)&&&&\\
 \hline
Marginal-Conj I&0.3&4&4685.7(84.104)&2382.799(169.359)\\
Marginal-Conj I&0.5&4&4757.2(37.077)&2944.065(195.011)\\
Marginal-Conj I&0.7&4&4655.2(52.514)& 2726.232(132.828)\\
Conditional-Conj I&0.5&-&10141.6(237.735)& 905.444(41.475)\\
 \hline
Normalized Stable process&&&&\\
 \hline
Marginal-Conj I&0.3&4&7658.3(193.773)&2630.264(429.877)\\
Marginal-Conj I&0.5&4&8203.1(106.798)&3139.412(351.788)\\
Marginal-Conj I&0.7&4&8095.7(85.2640)&2394.756(295.923)\\
Conditional-Conj I&0.5&-&10033.1(22.647)&912.382(167.089)\\
\hline
Normalized Generalized Gamma process ($ \tau = 1$)&&&&\\
 \hline
Marginal-Conj I&0.3&4&7685.8(208.98)&3587.733(569.984)\\
Marginal-Conj I&0.5&4&8055.6(93.164)&4443.905(367.297)\\
Marginal-Conj I&0.7&4&8117.9(113.188)&4936.649(411.568)\\
Conditional-Conj I&0.5&-&10046.9(206.538)&1000.214(70.148)\\
\hline
 \end{tabular}}
  \caption{\normalsize{Running times in seconds and number of cluster's ESS averaged over 5 chains.  Unidimensional dataset, 30,000 iterations per chain, 10,000 burn in. \label{assess}}}%
\end{table}

\spacingset{1.45}

 After assessing the algorithm's performance we used it for inference with a nonparametric mixture model where the top level is a prior from the $\sigma$-Stable Poisson-Kingman class. Since any prior in this class can be chosen, one possible criterion for model selection is predictive performance. In Table~\ref{predprobgalaxy}, we reported the average (leave one out) predictive probabilities, see the supplementary material for details. We can see that all priors in this class have similar average predictive probabilities but the NGG slightly outperforms the rest. 
 
  \begin{figure}[h]
    \vspace{-5mm}
\begin{centering}
\includegraphics[scale=0.35]{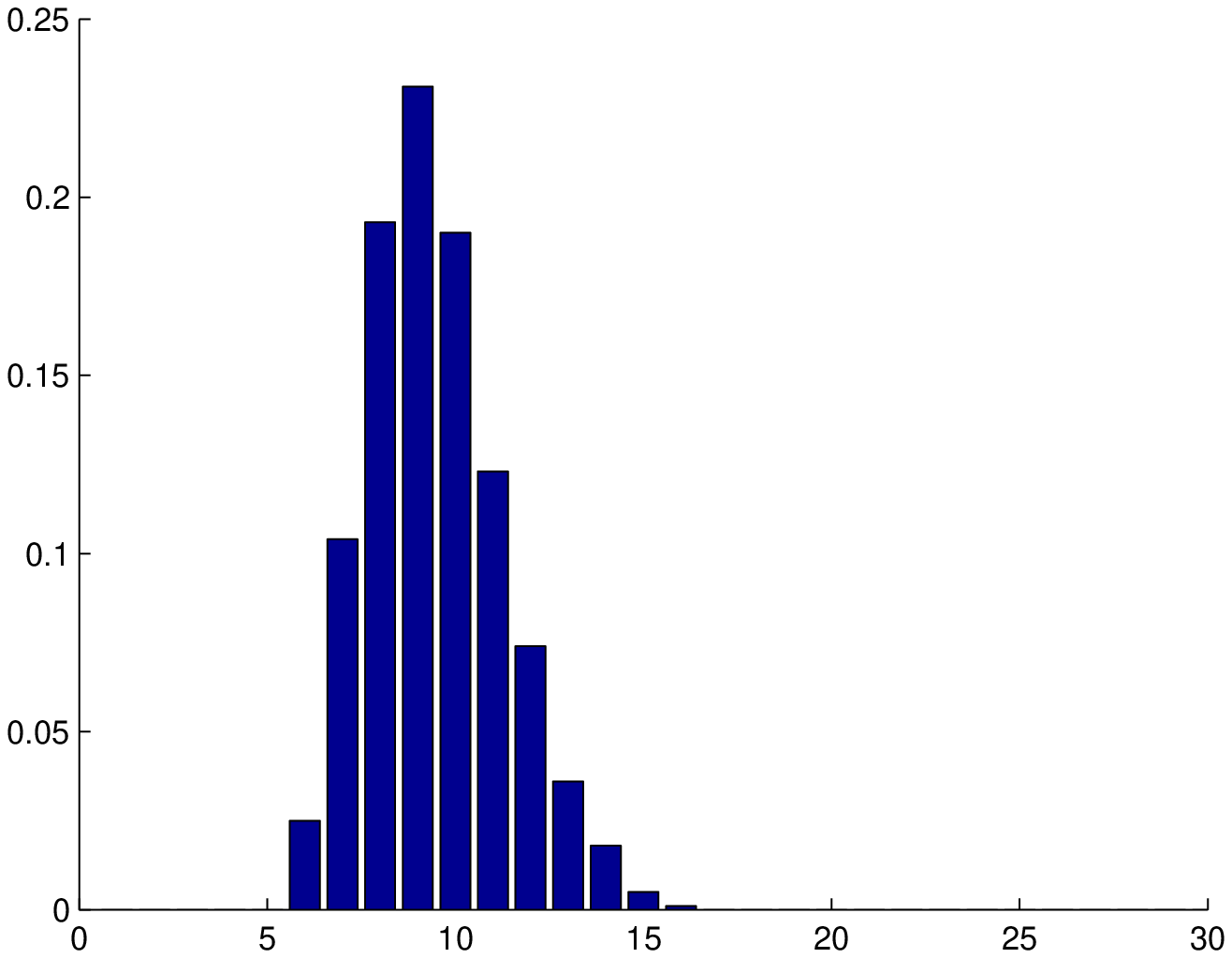}
\includegraphics[scale=0.25]{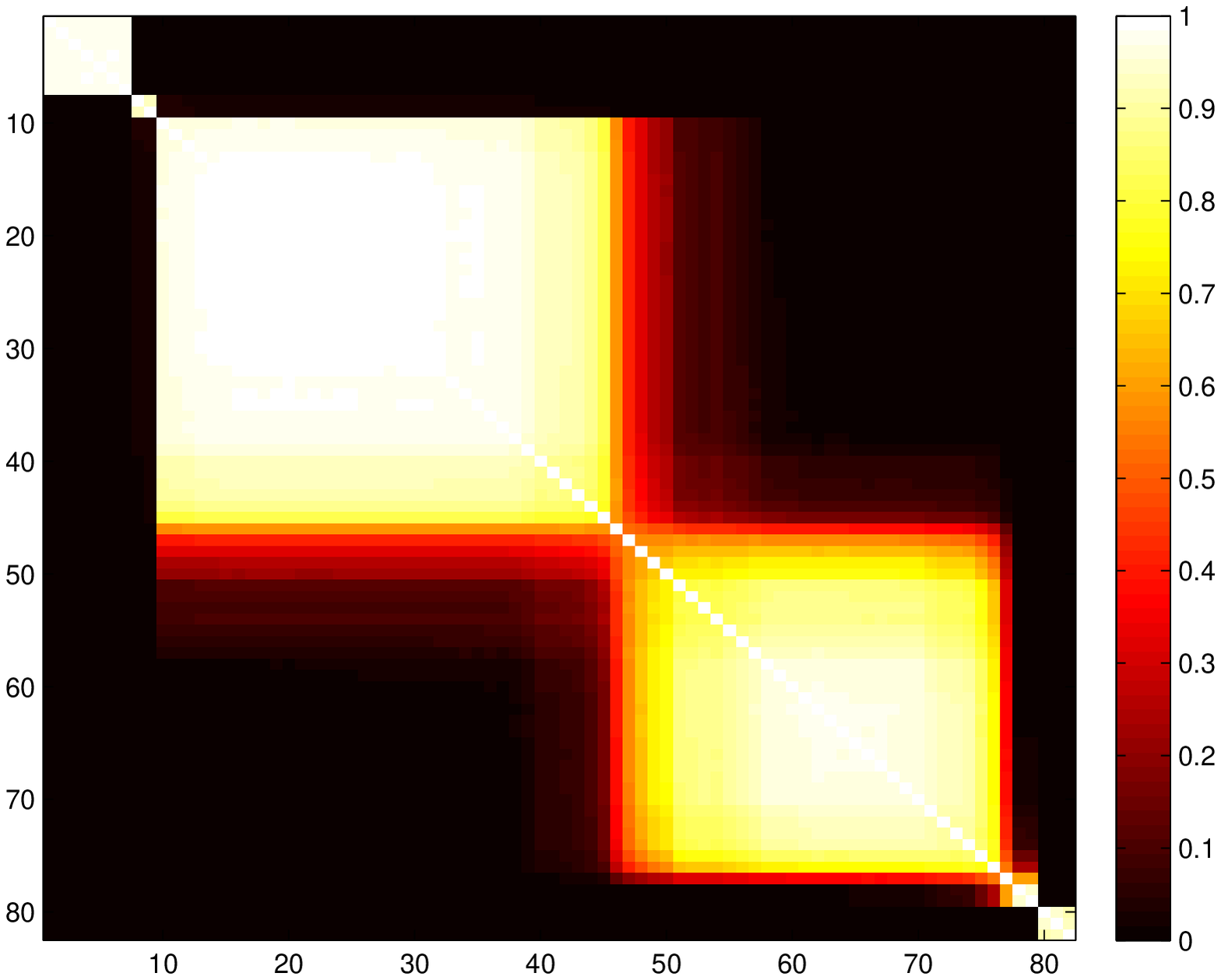}
\includegraphics[scale=0.4]{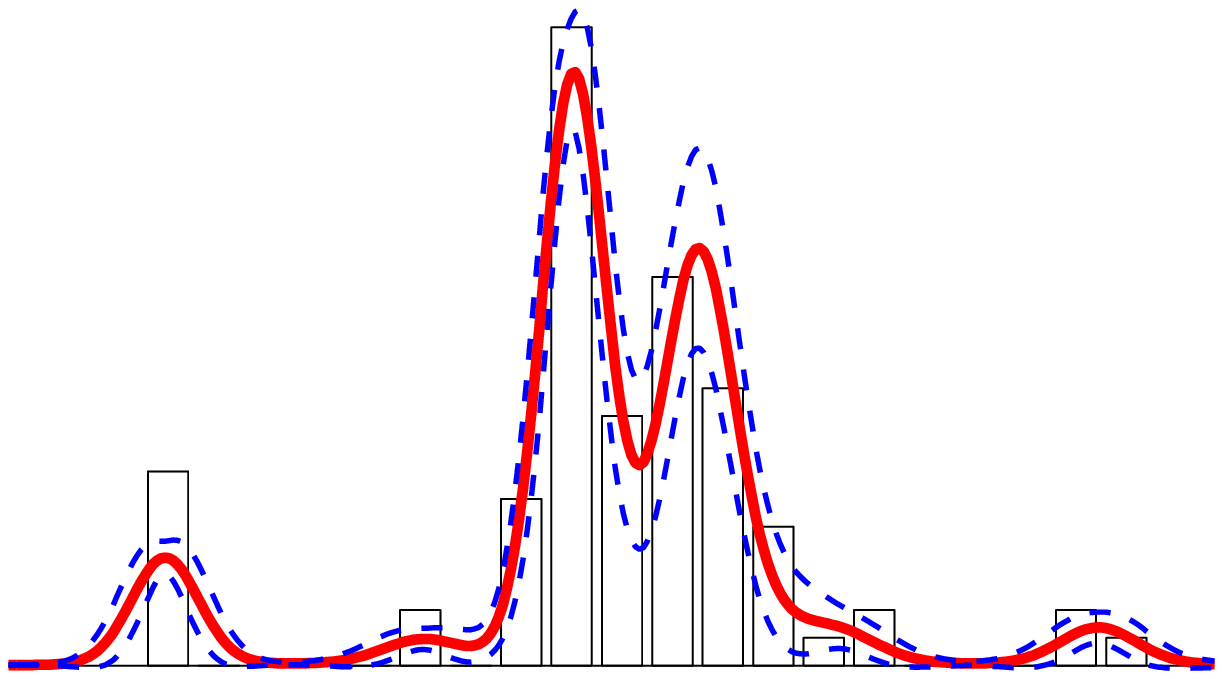}
\end{centering}\caption{\normalsize{210,000 iterations, 10,000 burn in and 20 thinning factor.}\label{galaxyfig}}
\end{figure}

\spacingset{1}
\begin{table}[h]
\centering
   \scalebox{0.8}{
\begin{tabular}{|c|c|}
\hline
Examples& Average Predictive Probability\\
\hline
PY&$0.1033(0.052)$\\
\textbf{NGG}&\textbf{ 0.1228(0.065)}\\
Gamma Tilted&$0.1186( 0.065)$\\\
NS&$ 0.1123( 0.057)$\\
\hline
\end{tabular}}\caption{Unidimensional experiment's average (leave one out) predictive probabilities.\label{predprobgalaxy}}
\end{table}

\spacingset{1.45}

In Figure~\ref{galaxyfig}, the mode of the posterior distribution is reported, it is around 10 clusters and there are 6 clusters in the coclustering probability matrix. Indeed, a good estimate of the density might include superfluous components having vanishingly small weights as explained in \citet{miller13}. The third plot shows the  corresponding density estimate which is consistent  under certain conditions as shown in \citet{DeBlasi2015}.

\subsection{Multidimensional experiment}

The dataset from \citet{mata11} consists of $n$ $D$-dimensional triacylglyceride profiles of different types of oils where $n=120$ and $D=4000$. The observations consist of profiles of olive, monovarietal vegetable and blends of oils. Within each type there could be several subtypes so we cannot know the number of varieties a priori.  We preprocessed the data by applying Principal Component Analysis (PCA) \citep{pca} to get the relevant dimensions in it, a useful technique when the signal to noise ratio is small. We used the first $d=8$ principal components which explained 97\% of the variance and encoded sufficient information for the mixture model to recover distinct clusters.

Then a $\sigma$-stable Poisson-Kingman mixture of multivariate Normals with unknown covariance matrix and mean vector was chosen for different $h$-tilting functions. A multivariate Normal-Inverse Wishart was chosen as a base measure and the corresponding likelihood $F$ for the $k$th cluster:
\begin{align}
H_0(\textrm{d}\mu_k,\textrm{d}\Sigma_k) = & \mathcal{N}_d\left(\textrm{d}\mu_k\mid\mu_0,r_0\Sigma_k^{-1}\right)\mathcal{IW}_d\left(\textrm{d}\Sigma_k\mid \nu_0, S_0\right)\nonumber\\
F(\textrm{d}x_1,\hdots,\textrm{d}x_{n_k}\mid \mu_k, \Sigma_k) = & \prod_{i=1}^{n_k}\mathcal{N}_d\left(\textrm{d}x_i\mid \mu_k, \Sigma_k\right)\nonumber
\end{align}
where $X_1,\hdots,X_{n_k}$ are the observations currently assigned to cluster $k$. $\mathcal{N}_d$ denotes a $d$-variate Normal distribution with given mean vector $\mu_k$ and covariance matrix $r_0\Sigma_k^{-1}$, $\mathcal{IW}_d$ denotes an inverse Wishart over $d\times d$ positive definite matrices with given degrees of freedom and scale matrix. The Inverse Wishart is parameterised as in \citet{GelCarSte1995a}. $S_0$ was chosen to be a diagonal matrix with each diagonal element given by the maximum range of the data across all dimensions and degrees of freedom $\nu = d+3$, a weakly informative case.

\begin{figure}[h!]
\noindent\begin{minipage}{\textwidth}

 \begin{minipage}[b]{.4\textwidth}
\includegraphics[scale=0.4]{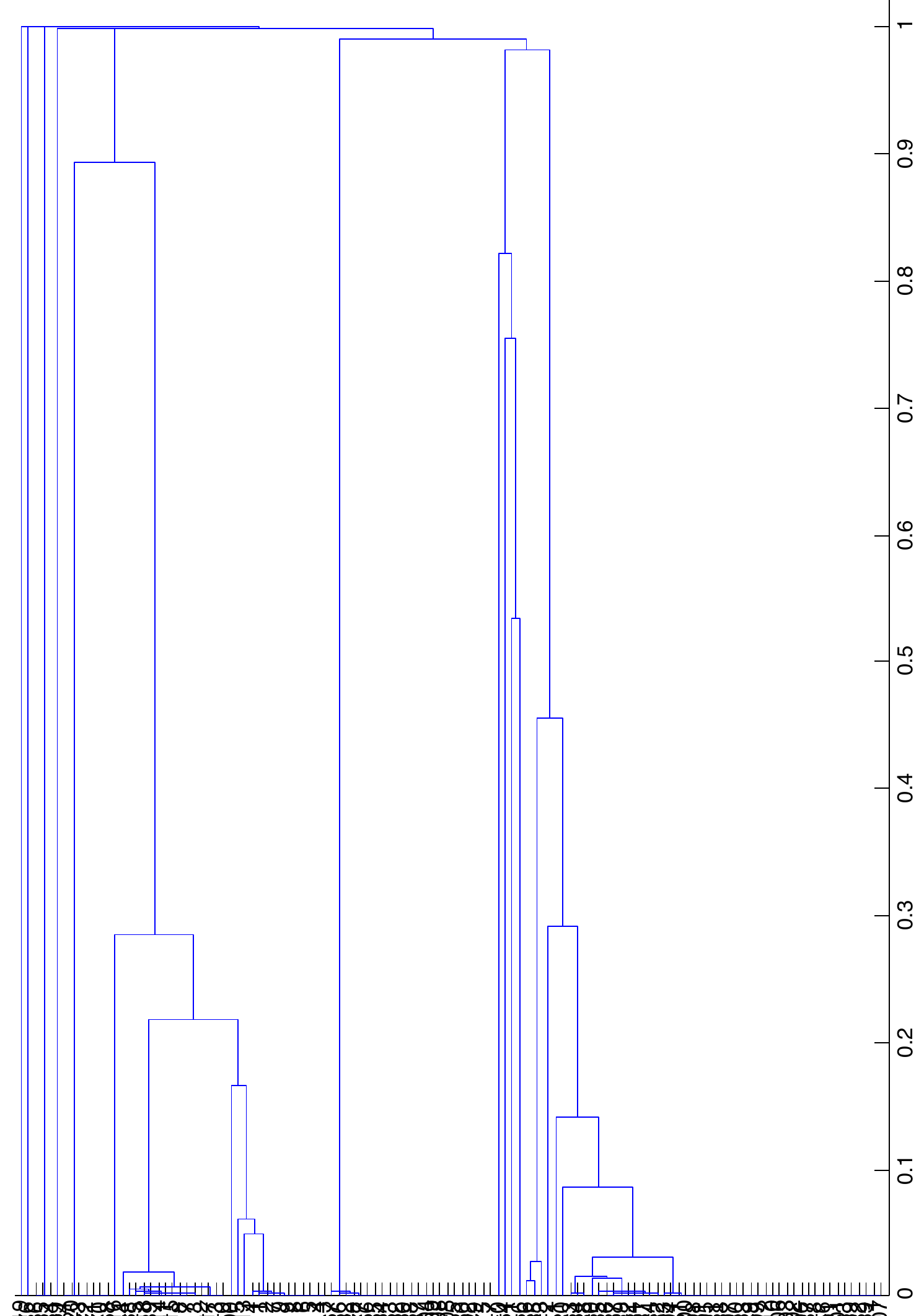}
\end{minipage}
\begin{minipage}[b]{.2\textwidth}

\includegraphics[scale=0.25]{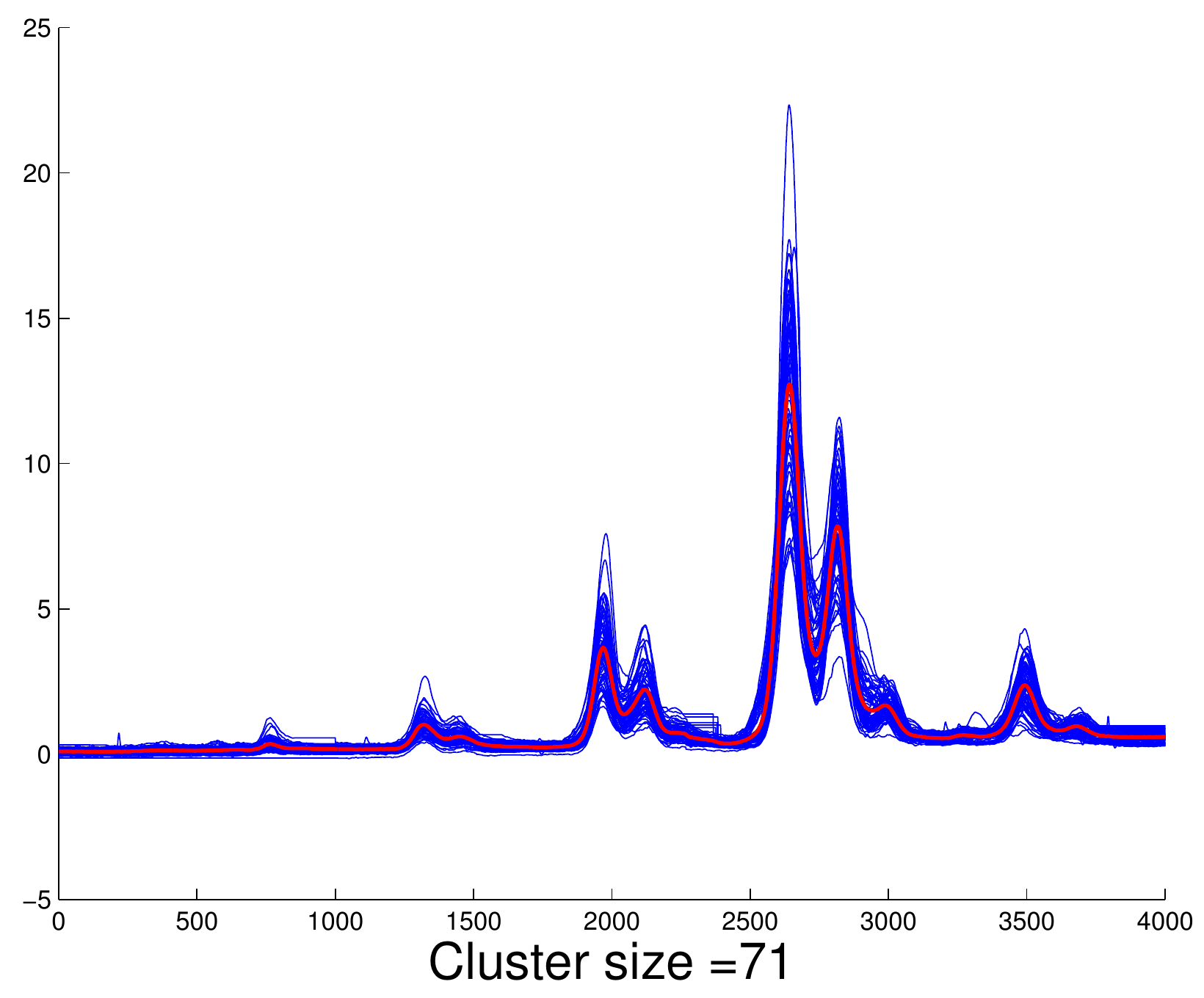}
\includegraphics[scale=0.25]{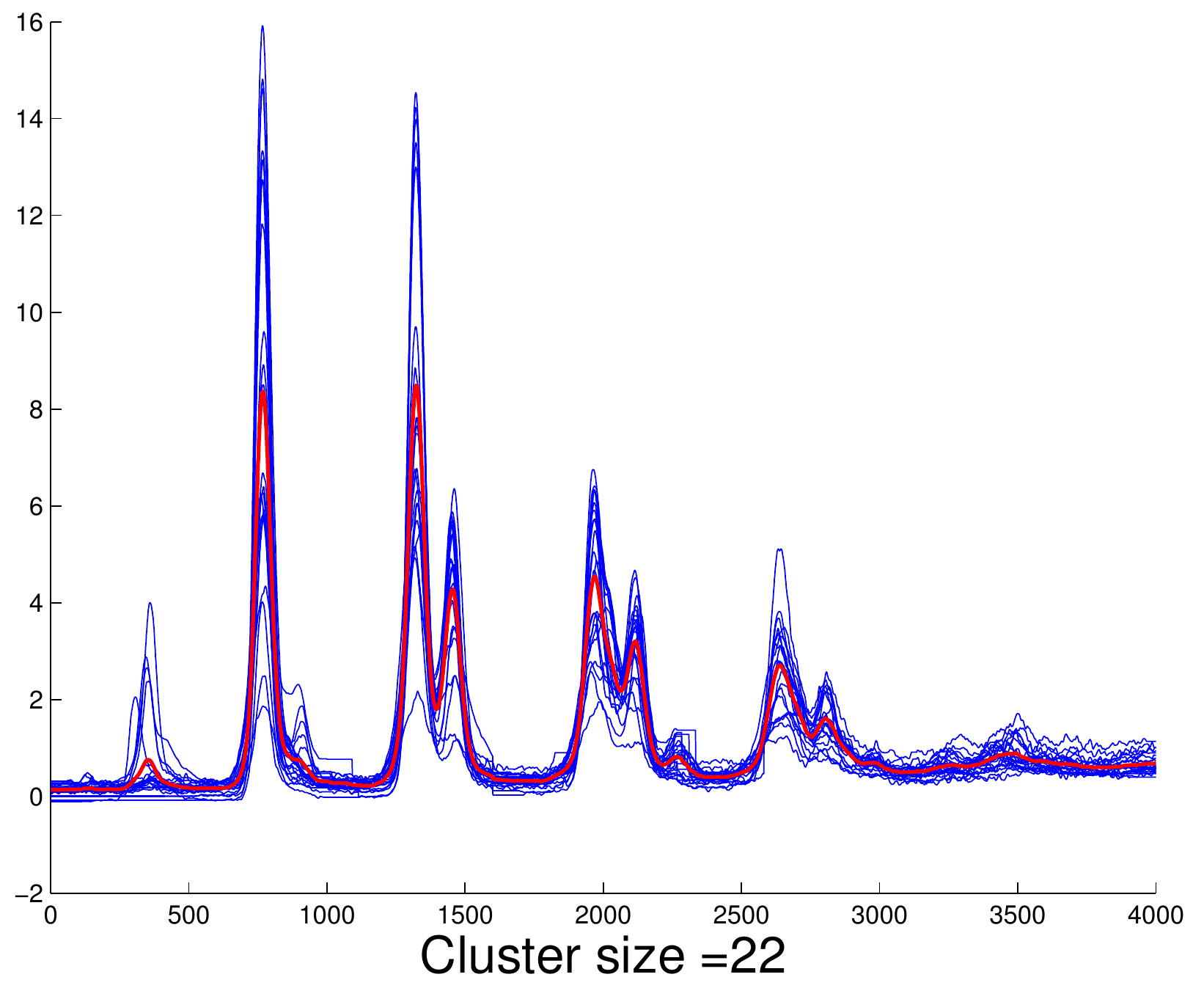}
\includegraphics[scale=0.25]{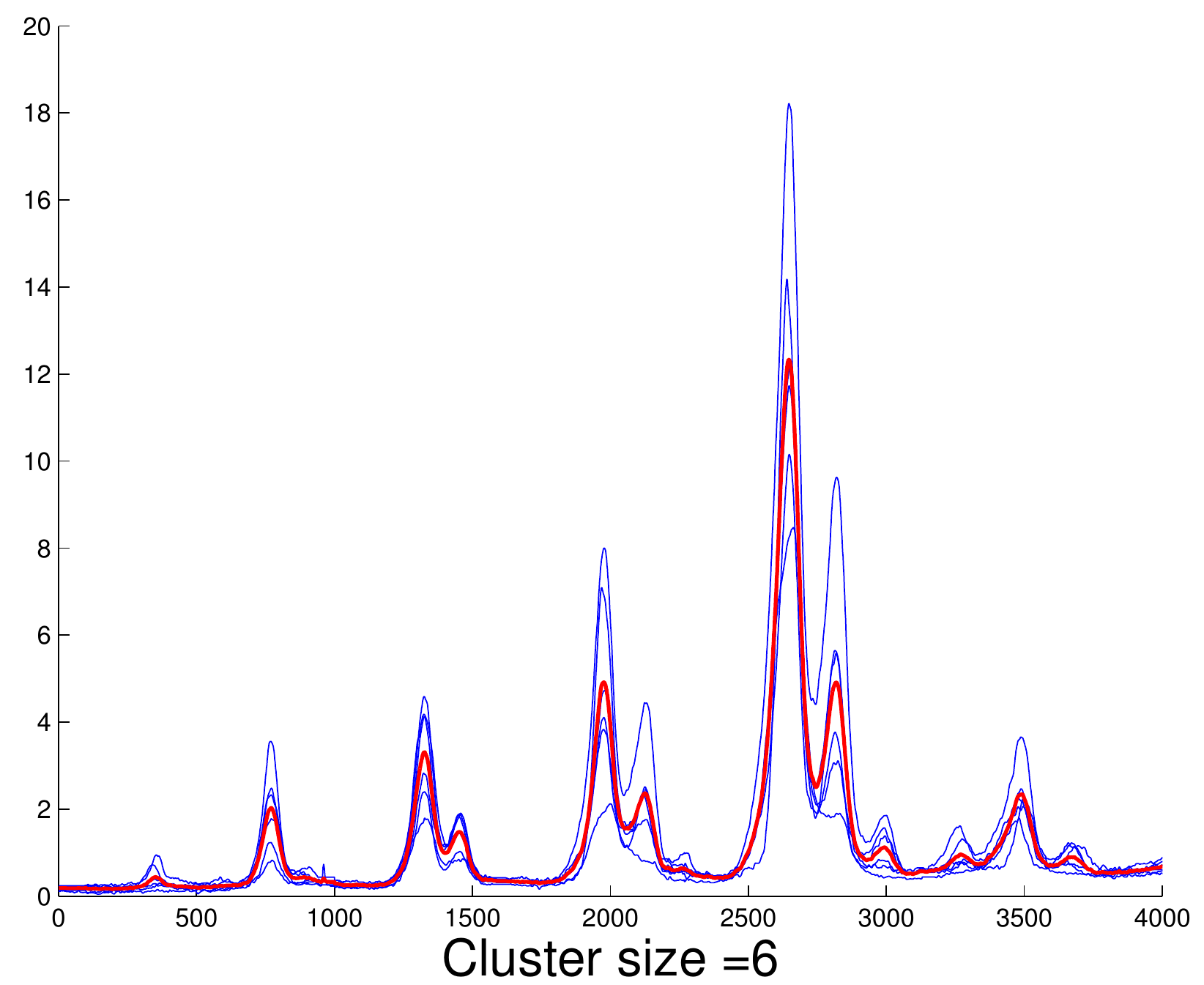}
\end{minipage}
\begin{minipage}[b]{.3\textwidth}
\includegraphics[scale=0.25]{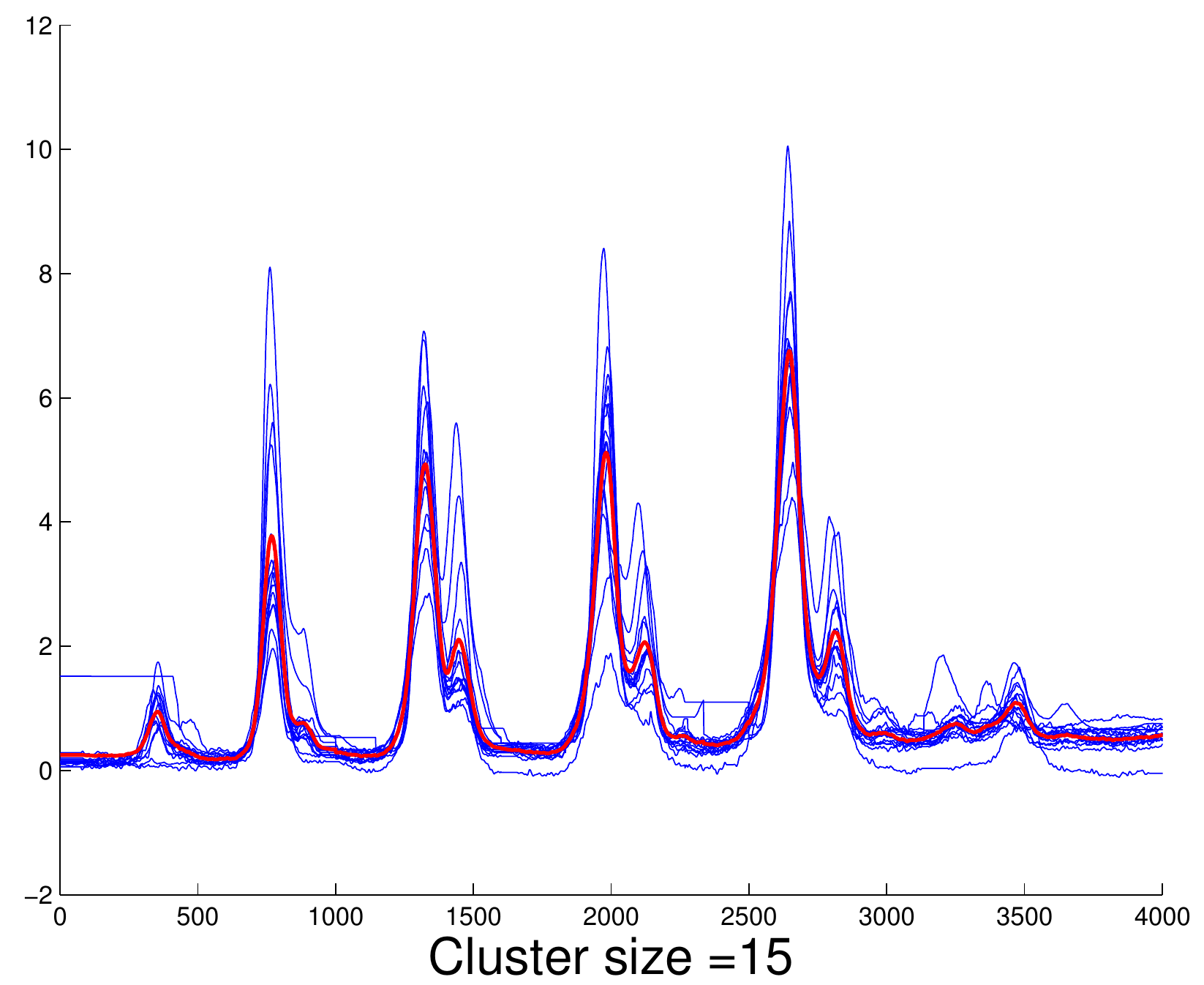}
\includegraphics[scale=0.25]{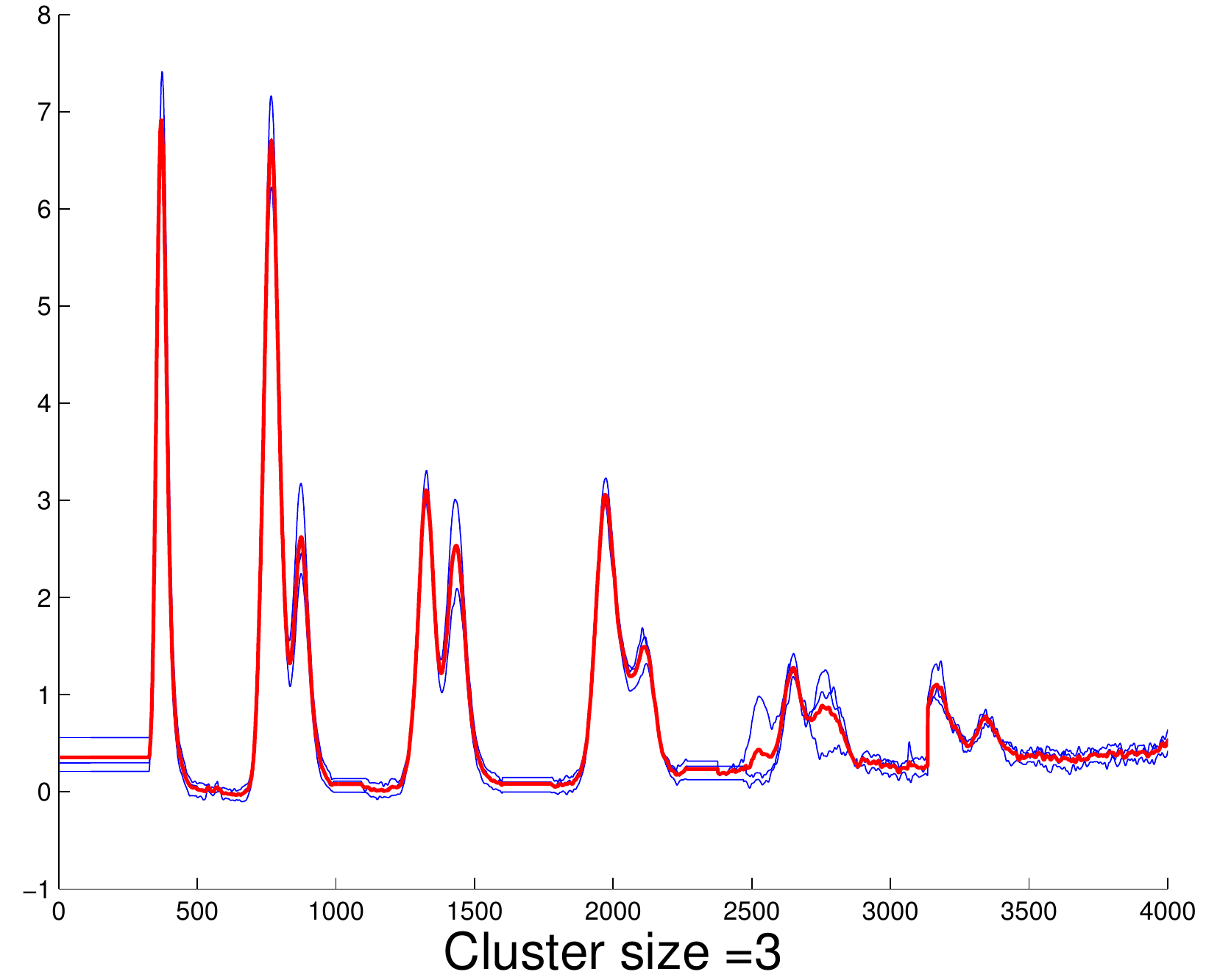}

\end{minipage}
\end{minipage}
\spacingset{1}
\caption{\normalsize{Dendrogram and mean profile per cluster (in red), profiles in each cluster (blue) using a $\NGG$ prior.} \label{oliveres}}
\end{figure}

 In Table~\ref{predprobolive}, the average (5-fold) predictive probabilities are reported, see supplementary material for details. Again, we observe that all priors in this class have similar average predictive probabilities. In Figure \ref{oliveres}, the mean curve per cluster and the coclustering probability matrix are reported. This mean curve reflects the average triacylglyceride profile per oil type. The coclustering probability matrix was used as an input to an agglomerative clustering algorithm to obtain a hierarchical clustering representation as in \citet{med02}. In certain contexts, it is useful to think of a hierarchical clustering  rather than a flat one, it might be natural to think of superclasses. %
 
\spacingset{1}

 \begin{table}
 \centering
 \scalebox{0.8}{
\begin{tabular}{|c|c|}
\hline
Examples& Average Predictive Probability (5 fold)\\
\hline
DP &5.5484e-12 (7.6848e-13)\\
PY&4.1285e-12 (7.5549e-13)\\
\textbf{NGG}& \textbf{9.6266e-12 (3.4035e-12)}\\
Gamma tilted&6.7099e-12(1.5767e-12)\\\
NS&8.3328e-12(9.7106e-13)\\
Lamperti tilted& 5.4251e-12 (1.0538e-12)\\
\hline
\end{tabular}}\caption{Multidimensional experiment's average (5-fold) predictive probabilities.\label{predprobolive}}
\end{table}

\spacingset{1.45}
 In Figure~\ref{oliveres}, the mean curves per cluster are shown. These were found by thresholding the hierarchy to 8 clusters and ignoring clusters of size one. The first plot corresponds to the olive oil cluster, it is well represented by the mean curve.  The last two plots correspond to data that belongs to non-olive blends of oil. The third and fourth plots correspond to non-olive monovarietal oil clusters. We could interpret the two clusters as different varieties of vegetable oil since their corresponding mean curves are indeed different. In the dendrogram it is clear that most of the data belongs to 3 large clusters and that $60\%$ of the triacylglycerides are olive oil.

\vspace{-5mm}

\spacingset{1.45}

\vspace{-5mm}

   \section{Discussion}

A completely random measure completely specifies its  total mass but if we allow the total mass to come from a different distribution, we obtain the class of Poisson-Kingman RPMs introduced by \citet{Pit2003a}.  If we restrict to the $\sigma$-stable L\'evy measure  we obtain the $\sigma$-stable Poisson Kingman class. This class  of random probability measures is natural but certain intractabilities have hindered its use. For instance, the intractability associated to the $\sigma$-stable density, as noted by \citet{lijoi08}. The aim of this paper was to review this class of priors and some characterisations which allowed us to build a novel algorithm for posterior simulation that is efficient and easy to use.

   Our algorithm is the first sampler of marginal type for mixture models with $\sigma$-stable Poisson-Kingman priors. Previously, a conditional sampler has been proposed by \citet{favaroslice12}. One of the advantages of our approach is that the number of auxiliary variables per iteration is smaller than the conditional sampler's, hence, it has smaller memory and storage requirements. It also has better ESS and running times as shown in our experiments. Both conditional and marginal samplers for this class are  general purpose: they do not depend on a particular characterisation of a specific Bayesian nonparametric prior as opposed to previous approaches. This makes them very useful and should be added to our Bayesian nonparametrics toolbox. Our approach could be used as a building block in a more complex model using the proposed algorithm. This is an interesting avenue of future research.

  \section{Acknowledgements}
Mar\'ia Lomel\'i thanks the Gatsby Charitable Foundation for generous funding. Stefano Favaro is supported by the European Research Council through StG N-BNP 306406. Yee Whye Teh is supported by the European Research Council under the European UnionÕs Seventh Framework Programme (FP7/2007-2013) ERC grant agreement no. 617411. 

\bibliographystyle{authordate1}
\bibliography{bibjcgs}

\begin{thebibliography}{}

\bibitem[\protect\citename{Aldous, }1985]{Ald1985a}
Aldous, D. 1985.
\newblock Exchangeability and Related Topics.
\newblock {\em Pages  1--198 of:} {\em {\'E}cole d'{\'E}t{\'e} de
  Probabilit{\'e}s de Saint-Flour XIII--1983}.
\newblock Springer, Berlin.

\bibitem[\protect\citename{Barrios {\em et~al.\ }\relax, }2013]{BarLijNie2013a}
Barrios, E., Lijoi, A., Nieto-Barajas, L.~E., \& Pr\"uenster, I. 2013.
\newblock Modeling with Normalized Random Measure mixture models.
\newblock {\em Statistical Science}, {\bf 28}, 283--464.

\bibitem[\protect\citename{Bertoin, }2006]{Ber2006a}
Bertoin, J. 2006.
\newblock {\em Random Fragmentation and Coagulation Processes}.
\newblock Cambridge University Press.

\bibitem[\protect\citename{Blackwell \& McQueen, }1973]{blackwell73}
Blackwell, D., \& McQueen, J.~B. 1973.
\newblock Ferguson distributions via {P\'o}lya urn schemes.
\newblock {\em Annals of Statistics}, {\bf 1}, 353--355.

\bibitem[\protect\citename{De~Blasi {\em et~al.\ }\relax, }2015]{DeBlasi2015}
De~Blasi, P., Favaro, S., Lijoi, A., Mena, R.~H., Pr\"uenster, I., \& Ruggiero,
  M. 2015.
\newblock Are Gibbs-type priors the most natural generalization of the
  {D}irichlet process?
\newblock {\em Pages  212--229 of:} {\em IEEE Transactions on Pattern Analysis
  \& Machine Intelligence},  vol. 37.

\bibitem[\protect\citename{de~la Mata-Espinosa {\em et~al.\ }\relax,
  }2011]{mata11}
de~la Mata-Espinosa, P., Bosque-Sendra, J.~M., Bro, R., \& Cuadros-Rodriguez,
  L. 2011.
\newblock Discriminating olive and Non-olive oils using {HPLC-CAD} and
  chemometrics.
\newblock {\em Analytical and Bioanalytical Chemistry}, {\bf 399}, 2083--2092.

\bibitem[\protect\citename{Devroye, }2009]{devroye}
Devroye, L. 2009.
\newblock Random variate generation for exponentially and polynomially tilted
  {S}table distributions.
\newblock {\em ACM Transactions on Modelling and Computer Simulation}, {\bf
  19}, 1--20.

\bibitem[\protect\citename{Escobar, }1994]{Esc1994a}
Escobar, M.~D. 1994.
\newblock Estimating normal means with a {D}irichlet process prior.
\newblock {\em Journal of the American Statistical Association}, {\bf 89},
  268--277.

\bibitem[\protect\citename{Escobar \& West, }1995]{EscWes1995a}
Escobar, M.~D., \& West, M. 1995.
\newblock {B}ayesian density estimation and inference using mixtures.
\newblock {\em Journal of the American Statistical Association}, {\bf 90},
  577--588.

\bibitem[\protect\citename{Ewens, }1972]{ewens72}
Ewens, W.~J. 1972.
\newblock The sampling theory of selectively neutral alleles.
\newblock {\em Theoretical Population Biology}, {\bf 3}, 3 87--112.

\bibitem[\protect\citename{Favaro \& Teh, }2013]{FavTeh2013a}
Favaro, S., \& Teh, Y.~W. 2013.
\newblock {MCMC} for Normalized Random Measure mixture models.
\newblock {\em Statistical Science}, {\bf 28}, 335--359.

\bibitem[\protect\citename{Favaro \& Walker, }2012]{favaroslice12}
Favaro, S., \& Walker, S.~G. 2012.
\newblock Slice sampling {$\sigma$-S}table {Poisson-Kingman} mixture models.
\newblock {\em Journal of Computational and Graphical Statistics}, {\bf 22},
  830--847.

\bibitem[\protect\citename{Favaro {\em et~al.\ }\relax, }2014]{FavLomTeh2014a}
Favaro, S., Lomeli, M., \& Teh, Y.~W. 2014.
\newblock On a class of {$\sigma$}-stable {P}oisson-{K}ingman models and an
  effective marginalized sampler.
\newblock {\em Statistics and Computing}, {\bf 25}, 67--78.

\bibitem[\protect\citename{Ferguson, }1973]{Fer1973a}
Ferguson, T.~S. 1973.
\newblock A {B}ayesian analysis of some nonparametric problems.
\newblock {\em Annals of Statistics}, {\bf 1}, 209--230.

\bibitem[\protect\citename{Gelman {\em et~al.\ }\relax, }1995]{GelCarSte1995a}
Gelman, A., Carlin, J., Stern, H., \& Rubin, D. 1995.
\newblock {\em {B}ayesian data analysis}.
\newblock Chapman \& Hall, London.

\bibitem[\protect\citename{Gnedin \& Pitman, }2006]{GnePit2006a}
Gnedin, A., \& Pitman, J. 2006.
\newblock Exchangeable {G}ibbs partitions and {S}tirling triangles.
\newblock {\em Journal of Mathematical Sciences}, {\bf 138}, 5674--5684.

\bibitem[\protect\citename{Griffin \& Walker, }2011]{GriWal2011a}
Griffin, J.~E., \& Walker, S.~G. 2011.
\newblock Posterior simulation of Normalized Random Measure mixtures.
\newblock {\em Journal of Computational and Graphical Statistics}, {\bf 20},
  241--259.

\bibitem[\protect\citename{Ho {\em et~al.\ }\relax, }2008]{james08}
Ho, M.~W., James, L.~F., \& Lau, J.~W. 2008.
\newblock {\em Gibbs partitions {(EPPFs)} derived from a {S}table subordinator
  are {F}ox {H} and {M}eijer {G} transforms}.
\newblock ArXiv:0708.0619.

\bibitem[\protect\citename{Ishwaran \& James, }2001]{IshJam2001a}
Ishwaran, H., \& James, L.~F. 2001.
\newblock {G}ibbs sampling methods for {Stick-Breaking} priors.
\newblock {\em Journal of the American Statistical Association}, {\bf 96},
  161--173.

\bibitem[\protect\citename{James, }2002]{Jam2002a}
James, L.~F. 2002.
\newblock {\em Poisson process partition calculus with applications to
  exchangeable models and {B}ayesian nonparametrics}.
\newblock ArXiv:math/0205093.

\bibitem[\protect\citename{James, }2013]{Jam2013a}
James, L.~F. 2013.
\newblock {\em {Stick-Breaking} {PG($\alpha,\psi$)-G}eneralized {G}amma
  processes}.
\newblock ArXiv:1308.6570.

\bibitem[\protect\citename{James {\em et~al.\ }\relax, }2009]{JamLijPru2009a}
James, L.~F., Lijoi, A., \& Pr\"uenster, I. 2009.
\newblock Posterior analysis for Normalized Random Measures with Independent
  Increments.
\newblock {\em Scandinavian Journal of Statistics}, {\bf 36}, 76--97.

\bibitem[\protect\citename{Jolliffe, }2002]{pca}
Jolliffe, I.T. 2002.
\newblock {\em Principal component analysis}. 2nd. edn.
\newblock Springer Verlag, Berlin.

\bibitem[\protect\citename{Kalli {\em et~al.\ }\relax, }2011]{kalliwalker11}
Kalli, M., Griffin, J.~E., \& Walker, S.~G. 2011.
\newblock {Slice sampling mixture models}.
\newblock {\em Statistics and Computing}, {\bf 21}, 93--105.

\bibitem[\protect\citename{Kanter, }1975]{Kan1975a}
Kanter, M. 1975.
\newblock Stable densities under change of scale and total variation
  inequalities.
\newblock {\em Annals of Probability}, {\bf 3}, 697--707.

\bibitem[\protect\citename{Kingman, }1967]{Kin1967a}
Kingman, J. F.~C. 1967.
\newblock Completely Random Measures.
\newblock {\em Pacific Journal of Mathematics}, {\bf 21}, 59--78.

\bibitem[\protect\citename{Kingman, }1975]{Kin1975a}
Kingman, J. F.~C. 1975.
\newblock Random discrete distributions.
\newblock {\em Journal of the Royal Statistical Society}, {\bf 37}, 1--22.

\bibitem[\protect\citename{Kingman, }1978]{Kin1978a}
Kingman, J. F.~C. 1978.
\newblock The representation of partition structures.
\newblock {\em Journal of the London Mathematical Society}, {\bf 18}, 374--380.

\bibitem[\protect\citename{Kingman, }1993]{Kin1993a}
Kingman, J. F.~C. 1993.
\newblock {\em {P}oisson processes}.
\newblock Oxford University Press.

\bibitem[\protect\citename{Laha \& Rohatgi, }1979]{Laha79}
Laha, R.~G., \& Rohatgi, V.K. 1979.
\newblock {\em Probability theory}.
\newblock John Wiley and Sons.

\bibitem[\protect\citename{Lijoi {\em et~al.\ }\relax, }2005]{LijMenPru2005a}
Lijoi, A., Mena, R.~H., \& Pr\"uenster, I. 2005.
\newblock Hierarchical mixture modelling with {N}ormalized {Inverse-G}aussian
  priors.
\newblock {\em Journal of the American Statistical Association}, {\bf 100},
  1278--1291.

\bibitem[\protect\citename{Lijoi {\em et~al.\ }\relax, }2007]{LijMenPru2007a}
Lijoi, A., Mena, R.~H., \& Pr\"uenster, I. 2007.
\newblock Controlling the reinforcement in {B}ayesian nonparametric mixture
  models.
\newblock {\em Journal of the Royal Statistical Society B}, {\bf 69}, 715--740.

\bibitem[\protect\citename{Lijoi {\em et~al.\ }\relax, }2008]{lijoi08}
Lijoi, A., Pr\"{u}nster, I., \& Walker, S.~G. 2008.
\newblock Investigating nonparametric priors with {G}ibbs structure.
\newblock {\em Statistica Sinica}, {\bf 18}, 1653 -- 1668.

\bibitem[\protect\citename{Lo, }1984]{Lo1984a}
Lo, A.Y. 1984.
\newblock On a class of Bayesian nonparametric estimates: I. density estimates.
\newblock {\em Annals of Statistics}, {\bf 12}, 351--357.

\bibitem[\protect\citename{MacEachern, }1994]{Mac1994a}
MacEachern, S.~N. 1994.
\newblock Estimating {N}ormal means with a conjugate Style {D}irichlet process
  prior.
\newblock {\em Communications in Statistics: Simulation and Computation}, {\bf
  23}, 727--741.

\bibitem[\protect\citename{Medvedovic \& Sivaganesan, }2002]{med02}
Medvedovic, M., \& Sivaganesan, S. 2002.
\newblock Bayesian infinite mixture model based clustering of gene expression
  profiles.
\newblock {\em Bioinformatics}, {\bf 18}, 1194--1206.

\bibitem[\protect\citename{Miller \& Harrison, }2013]{miller13}
Miller, J.~W., \& Harrison, T.~M. 2013.
\newblock A simple example of {D}irichlet process mixture inconsistency for the
  number of components.
\newblock {\em In:} {\em Neural Information Processing Systems}.

\bibitem[\protect\citename{Neal, }2000]{Nea2000a}
Neal, R.~M. 2000.
\newblock {M}arkov chain sampling methods for {D}irichlet process mixture
  models.
\newblock {\em Journal of Computational and Graphical Statistics}, {\bf 9},
  249--265.

\bibitem[\protect\citename{Neal, }2003]{Nea2003a}
Neal, R.~M. 2003.
\newblock Slice sampling.
\newblock {\em Annals of Statistics}, {\bf 31}, 705--767.

\bibitem[\protect\citename{Nieto-Barajas {\em et~al.\ }\relax,
  }2004]{NiePruWal2004a}
Nieto-Barajas, L.~E., Pruenster, I., \& Walker, S.~G. 2004.
\newblock Normalized Random Measures driven by increasing additive processes.
\newblock {\em Annals of Statistics}, {\bf 32}, 2343--2360.

\bibitem[\protect\citename{Papaspiliopoulos \& Roberts, }2008]{PapRob2008a}
Papaspiliopoulos, O., \& Roberts, G.~O. 2008.
\newblock {Retrospective Markov chain Monte Carlo methods for Dirichlet process
  hierarchical models}.
\newblock {\em Biometrika}, {\bf 95}, 169--186.

\bibitem[\protect\citename{Perman {\em et~al.\ }\relax, }1992]{PerPitYor1992a}
Perman, M., Pitman, J., \& Yor, M. 1992.
\newblock Size-biased sampling of {P}oisson point processes and excursions.
\newblock {\em Probability Theory and Related Fields}, {\bf 92}, 21--39.

\bibitem[\protect\citename{Pitman, }2003]{Pit2003a}
Pitman, J. 2003.
\newblock {P}oisson-{K}ingman partitions.
\newblock {\em Pages  1--34 of:} Goldstein, D.~R. (ed), {\em Statistics and
  Science: a Festschrift for Terry Speed}.
\newblock Institute of Mathematical Statistics.

\bibitem[\protect\citename{Pitman, }2006]{Pit2006a}
Pitman, J. 2006.
\newblock {\em Combinatorial Stochastic Processes}.
\newblock Lecture Notes in Mathematics.
\newblock Springer-Verlag, Berlin.

\bibitem[\protect\citename{Pitman \& Yor, }1997]{PitYor1997a}
Pitman, J., \& Yor, M. 1997.
\newblock The two parameter {P}oisson-{D}irichlet distribution derived from a
  {S}table subordinator.
\newblock {\em Annals of Probability}, {\bf 25}, 855--900.

\bibitem[\protect\citename{Regazzini {\em et~al.\ }\relax,
  }2003]{RegLijPru2003a}
Regazzini, E., Lijoi, A., \& Pr\"uenster, I. 2003.
\newblock Distributional results for means of random measures with independent
  increments.
\newblock {\em Annals of Statistics}, {\bf 31}, 560--585.

\bibitem[\protect\citename{Roeder, }1990]{galaxy}
Roeder, K. 1990.
\newblock Density estimation with confidence sets exemplified by super-clusters
  and voids in the galaxies.
\newblock {\em Journal of the American Statistical Association}, {\bf 85},
  617--624.

\bibitem[\protect\citename{Tanner \& Wong, }1987]{tanner87}
Tanner, M., \& Wong, W. 1987.
\newblock The calculation of posterior distributions by data augmentation.
\newblock {\em Journal of the American Statistical Association}, {\bf 82},
  528--550.

\bibitem[\protect\citename{Walker, }2007]{Wal2007a}
Walker, S.~G. 2007.
\newblock Sampling the {D}irichlet mixture model with slices.
\newblock {\em Communications in Statistics - Simulation and Computation}, {\bf
  36}, 45--54.

\bibitem[\protect\citename{Wuertz {\em et~al.\ }\relax, }2013]{WueMaeRme2013a}
Wuertz, A., Maechler, M., \& {Rmetrics core team members}. 2013.
\newblock {\em stabledist}.
\newblock CRAN R Package Documentation.

\bibitem[\protect\citename{Zolotarev, }1966]{zolotarev66}
Zolotarev, V.~M. 1966.
\newblock On the representation of {S}table laws by integrals.
\newblock {\em Selected Translations Math Statist. and Probability}, {\bf 6},
  84--88.

\end{thebibliography}

\spacingset{1.45}
\appendix
\section{Matlab Code}
\begin{description}

\item[Matlabcode\_Marginalsampler:] The folder contains Matlab files with code to perform posterior inference as described in the article for a unidimensional and multidimensional datasets. The folder also contains all datasets used as examples in the articles and routines for creating plots. (.zip file)

\item[Galaxy data set:] Data set from   \cite{galaxy} used in the unidimensional illustration in Section~ 6.1. (.txt file)
\item[Olive oil data set:] Data set   \cite{mata11} used in the multidimensional illustration in Section~ 6.2. (.mat file)

\end{description}
\section{Additional pseudocode}
\spacingset{1}
\begin{algorithm}[H]
\caption{SliceSampler($x_j,f,E,L$)}
\begin{algorithmic}
\State $N_1=N_2=0$
\State Sample $y\sim \textrm{U}\left[0, f(x_j)\right]$\Comment $f$ can be an unnormalized density
\State Sample $l\sim \textrm{U}(0,L)$ \Comment  $L$ is the chosen length of the interval $(a,b)$
\State Set $a=x_j-l$, $b=x_j-l+L$\Comment $x_j$ is the previously accepted point 
\While{$f\left( a\right)<y\vee f\left(b \right)<y$}\Comment Samples $x_{j+1}$ uniformly from the set $f^{\left[-1\right]}[y,\infty)$
\If {$f\left(a \right)<y$}
 \State $a = a-E$\Comment $E$ is the chosen size to enlarge the initial interval
 \State $N_1=N_1+1$
\EndIf
\If {$f\left(b \right)<y$}
 \State $b = b+E$
  \State $N_2=N_2+1$
\EndIf
\State $L=b-a$
\State Sample $l\sim \textrm{U}(0,L)$
\State Set $a=x_j-l$, $b=x_j-l+L$
\EndWhile 
\State Set $c=x_j-l-N_1L$, $d =x_j-l+N_2L$
\State Sample $w\sim \textrm{U}(c,d)$
\While{$f(w)<y$}
\If {$w<x_j<b$}
\State $w\sim \textrm{U}(w,d)$
\Else
\State $w\sim \textrm{U}(c,w)$
\EndIf
\EndWhile
\State Return $x_{j+1}=w$

\end{algorithmic}
\end{algorithm}
\vspace{-4mm}
\begin{algorithm}\caption{CoClustering$\left(C\right)$}%
\begin{algorithmic}

\For {$m = 1 \to M, i=1 \to n, j=1 \to n$} \Comment $M$ is the number of MCMC iterations
 \If {$c_{m,i} == c_{m,j}$}\Comment $n$ is the number of data points
 \State $A_{i,j,m} =1$\Comment $A$ is a $n\times n \times M$ array
 \Else 
 \State  $A_{i,j,m} =0$
\EndIf
\EndFor
\State $P =\textrm{Sum}(A,3)/M$
\State \Return
\end{algorithmic}
\end{algorithm}

\medskip

\spacingset{1.45}

\section{Average leave-one-out and 5-fold predictiverobabilities}
The posterior predictive density for $X_{n+1}$ given $X$, $P$ and $\textbf{Y}=\left\{Y_1^*,\hdots,Y_k^*\right\}$ is

\begin{align*}
f_{\sigma}\left[X_{n+1} \in \textrm{d}x \mid R,W, \textbf{X},\textbf{Y}^*\right] &=
\eta_0(n+1,\sigma, k, w, r)\int{f\left(x_{n+1}\mid y\right)p(y)\textrm{d}y}\\&+
\eta_1(n+1,k)\sum_{j=1}^k{(n_j-\sigma)f\left(x_{n+1}\mid y^*_j\right)}\\
\intertext{where} 
\eta_0(n+1,\sigma, k, w, r)&=\frac{e^{-w(1-\sigma)}r^{-\sigma} }{\Gamma \left(n+1-\sigma(k+1)\right)}\\
\eta_1(\sigma,k)&=\frac{1}{\Gamma \left(n+1-\sigma k\right)}
\end{align*}
The corresponding empirical estimator, where $M$ is the size of the chain after burn in, is given by:
\begin{align*}
\hat{f}_{\sigma}\left[X_{n+1} \in \textrm{d}x \mid R,W, \textbf{X},\textbf{Y}^*\right] &=
\frac{1}{M} \left[\sum_{m=1}^M
\eta_0(n+1,\sigma, k_m, w_m, r_m)\int{f\left(x_{n+1}\mid y\right)p(y)\textrm{d}y}\right]
\\
&+\left[
\eta_1(n+1,k _m)\sum_{j=1}^{k_m}{(n_{j,m}-\sigma)f\left(x_{n+1}\mid y^{*}_{j,m} \right)}\right]
\end{align*}
To  obtain the average leave-one-out predictive probability we do the following:
\begin{enumerate}
\item  For $i=1,\hdots,n$, remove the $i$th observation from the dataset and run the MCMC with the remaining data points.
\item At each MCMC iteration evaluate the predictive probability of the $i$th datapoint .
\item Average over the $M$ MCMC iterations to get the average predictive probability for the $i$th observation.
\end{enumerate}
After we do this for all observations we take the average of their corresponding predictive probabilities to obtain the average leave one out predictive probability reported in Table 2.
\\
To  obtain the average 5-fold predictive probability we do the following:
\begin{enumerate}
\item  For $j=1,\hdots,5$ Randomly split the dataset into  training data (4/5) and test data (1/5) (making sure each observation belongs to the test data only once, in other words, sample without replacement).
.\item At each MCMC iteration evaluate the predictive probability of each test data point and take the average.
\item Average over the $M$ MCMC iterations to get the average predictive probability for the $j$th batch of test data.
\end{enumerate}
After we do this for all test data batches we take the average of their corresponding predictive probabilities to obtain the average 5-fold predictive probability reported in Table 3.
\section{Examples of EPPFs obtained from Proposition 1}
\begin{example}[\cite{JamLijPru2009a}]\label{nrmeppf}
The EPPF of the exchangeable partition $\Pi$ induced by $\NRM(\rho,H_0)$ is given by:
\begin{align}
\PP_{\rho,H_0}(\Pi_n=(c_k)_{k\in[K]}) &= \int_{\mathbb{R}^+} \frac{u^{n-1}}{\Gamma(n)} e^{-\psi_\rho(u)}du\prod_{k=1}^{K} \kappa_\rho(|c_k|,u)\nonumber\\
\intertext{where}
\psi_\rho(u) &= -\log \int_{\mathbb{R}^+} e^{-ut} f_\rho(t) dt = \int_{\mathbb{R}^+}(1-e^{-us})\rho(ds)\nonumber\\
\kappa_\rho(m,u)&= \int_{\mathbb{R}^+} v^{m}e^{-uv}\rho(dv).\nonumber
\end{align}
\end{example}
Example ~\ref{nrmeppf} can be obtained from Proposition 1 by introducing a disintegration
\begin{align}
T^{-n} = \int_{\mathbb{R}^+} \frac{u^{n-1}e^{-uT}}{\Gamma(n)} du\nonumber,
\end{align}
performing a change of variable $S=T-\sum_{k=1}^{K} V_k$, and marginalizing out $S$ and $(V_k)_{k\in[K]}$.

\begin{example}\label{nggeppf}The EPPF of the exchangeable partition $\Pi$ induced by $\NGG(\sigma, \tau)$ is given by:
\begin{align}
\PP_{\rho,H_0}(\Pi_n=(c_k)_{k\in[K]}) &= \int_{\mathbb{R}^+} \frac{u^{n-1}}{\Gamma(n)} e^{-\psi_\rho(u)}du\prod_{k=1}^{K} \kappa_\rho(|c_k|,u)\nonumber\\
\intertext{where}
\psi_\rho(u) &= -\log \int_{\mathbb{R}^+} e^{-ut} f_\rho(t) dt = \int_{\mathbb{R}^+}(1-e^{-us})\rho(ds)\nonumber\\
& = \int_{\mathbb{R}^+}(1-e^{-us})\frac{a}{\Gamma(1-\sigma)}s^{-\sigma-1}e{-\tau s}\textrm{d}s\nonumber
\end{align}
\begin{align}
\kappa_\rho(m,u)&= \int_{\mathbb{R}^+} v^{m}e^{-uv}\rho(dv)\nonumber\\
&= \frac{a}{\Gamma(1-\sigma)}\frac{\Gamma(m-\sigma)}{(u+\tau)^{m-\sigma}}\nonumber.
\end{align}
\end{example}
Example~\ref{nggeppf} can be obtained if we plug in the $\NGG$'s L\'evy measure in Example ~\ref{nrmeppf}.
\begin{example}\label{pyeppf}
The EPPF of the exchangeable partition $\Pi$ induced by PY$(\theta,\sigma)$ is given by:
\begin{align}
\mathbb{P}\left(\Pi_n=(c_k)_{k\in[K]}\right) =&\sigma^{K}\frac{\Gamma(\theta)}{\Gamma(\frac{\theta}{\sigma})}\int_{\mathbb{R}^+} \frac{r^{\theta+K\sigma-1}\exp{(-r^{\sigma})}}{\Gamma(\theta+n)} \prod_{k=1}^{K} \frac{\Gamma(|c_k|-\sigma)}{\Gamma(1-\sigma)}\nonumber\\
=&\sigma^{K}\frac{\Gamma(\theta)}{\Gamma(\frac{\theta}{\sigma})}\frac{\Gamma(\frac{\theta}{\sigma}+K)}{\Gamma(\theta+n)}\prod_{k=1}^{K} \frac{\Gamma(|c_k|-\sigma)}{\Gamma(1-\sigma)}\nonumber.
\end{align}
\end{example}
The EPPF can be obtained from Proposition 1 by introducing a disintegration 
\begin{align}
T^{-(n+\theta)} &= \int_{\mathbb{R}^+} \frac{r^{n+\theta-1}e^{-rT}}{\Gamma(n+\theta)} dr\nonumber,
\end{align}
performing a change of variables $S=T-\sum_{k=1}^{K} V_k$, $W=R^{\sigma}$ and marginalizing out $S$ and $(V_k)_{k\in[K]}$.
\begin{example}\label{crpeppf}
The EPPF of the exchangeable partition $\Pi$ induced by the DP$(\theta)$ with concentration parameter $\theta$ is given by:
\begin{align}
\mathbb{P}\left(\Pi_n=(c_k)_{k\in[K]}\right) =&\int_0^{\infty}\int_{\mathcal{V}_K} t^{-n}f_\rho(t-\textstyle\sum_{k=1}^{K} v_k) dt
\displaystyle \prod_{k=1}^{K} v_k^{|c_k|} \rho(dv_k)\nonumber \\
=&\frac{\theta^K\Gamma(\theta)}{\Gamma(n+\theta)}\prod_{k=1}^K{\Gamma(|c_k|)} \nonumber
\intertext{where}
\rho(x) &= \theta x^{-1}\exp{\left(-x\right)}\nonumber\\
f_{\rho}(x)&=\frac{1}{\Gamma(\theta)}x^{\theta-1}\exp{\left(-x\right)}.\nonumber
\end{align}
are the L\'evy measure of the Gamma Process and the corresponding density function, respectively.
\end{example}
Example ~\ref{crpeppf} can be obtained from Proposition 1 by performing a change of variables $P_k=V_k/T$ for $k=1,\hdots,K-1$, and marginalizing out $T$ and $(P_k)_{k\in[K]}$.\\

\section{Proof of Proposition in \cite{PerPitYor1992a}}
\begin{proposition*}[\cite{PerPitYor1992a}]\label{perman}
The sequence of surplus masses $(S_k)_{k\ge 0}$ forms a Markov chain, with initial distribution and transition kernels:
\begin{align}
\PP_{\rho,H_0}(S_0\in ds_0) &= f_\rho(s_0)ds_0\nonumber \\
\PP_{\rho,H_0}(S_k\in ds_k|S_1\in ds_1,\ldots,S_{k-1}\in ds_{k-1}) &=
\PP_{\rho,H_0}(S_k\in ds_k|S_{k-1}\in ds_{k-1}) \nonumber \\
&= \frac{(s_k-s_{k-1})\rho(d(s_k-s_{k-1}))}{s_{k-1}}\frac{f_\rho(s_k)}{f_\rho(s_{k-1})}.  \nonumber
\end{align}
\end{proposition*}
The proposition can be proven by induction, with each step being an application of the Palm formula (\cite{Kin1993a}) for Poisson processes, similar to the proof of Proposition 2.4 in \cite{Ber2006a}, as follows:
\begin{proof}

 \textbf{Induction over $k$, where $k$ is the number of size biased picks.}
  Let $\mu=\sum_{k=1}^{\infty}{\omega_k\delta_{x_k}}$ be an homogeneous completely random measure.
\begin{enumerate}

\item \textbf{Case k = 1}. A size biased sample can be obtained from it in the following way: we first pick the $k$th atom with probability $\frac{\omega_k}{T}$ where $T=\sum_{k=1}^{\infty}{\omega_k}$, the $kth$ surplus is $S_k=T-\sum_{j=1}^{k}{W_j}$ and set $W_1^*=\omega_k$. We are interested in the following conditional expectation:
\begin{align}
\mathbb{E}\left(W_1^*\in \textrm{d}\omega \mid T\in\left[t,t+\epsilon\right]\right)&= \frac{\mathbb{E}\left(W_1^*\in \textrm{d}\omega,T\in \left[t,t+\epsilon\right]\right) }{\mathbb{E}\left(T\in \left[t,t+\epsilon\right]\right) }\label{condexp1}
\end{align}
since we can formally define the conditional probability of interest as a weak limit of the conditional expectations indexed by $\epsilon$. Furthermore, in this proof it is only done for a simple function of the form $\delta_{W_1}(\textrm{d}\omega)$ but it can be easily extended for a measurable function $f$. We can proceed in the following way: if we consider a general simple function, then a non-negative measurable function as a limit of such simple functions and then decompose a measurable function $f$ in its negative an positive parts (see  \cite{Laha79}).
If we draw from the random measure
\begin{align}
\PP\left(W_1^*\in \textrm{d}\omega,T\in \textrm{d}t\mid \mu\right) &= \sum_{k=1}^{\infty}{\frac{\omega_k}{T}}\delta_{\omega_k}(\textrm{d}\omega)\delta_{T}(\textrm{d}t)\nonumber\\
\intertext{and then average over it we obtain the numerator in ~\eqref{condexp1}:}
\mathbb{E}\left(W_1^*\in \textrm{d}\omega,T\in \left[t,t+\epsilon\right]\right) &= \mathbb{E}\left(\sum_{k=1}^{\infty}{\frac{\omega_k}{T}}\delta_{\omega_k}(\textrm{d}\omega),T\in\left[t,t+\epsilon\right])\right)\nonumber\\
& = \int{\rho(y)y\delta_y(dw)\mathbb{E}\left((y+T)^{-1},y+T \in \left[t,t+\epsilon\right]\right)}\nonumber\\
& \rightarrow \frac{\epsilon}{t}\rho(\textrm{d}\omega)\omega f_{\rho}(t-\omega)\hspace{4mm}\textrm{as} \hspace{4mm}\epsilon\rightarrow 0+\nonumber
\end{align} 
For the third equality, we use Palm's Formula where $G(M,f)=\frac{1}{T}\delta_{y}(\textrm{d}\omega)\mathbb{I}_{\left[t,t+\epsilon\right]}(T)$ and for the fourth
\begin{align}
\mathbb{E}\left((y+T)^{-1},y+T \in \left[t,t+\epsilon\right]\right) &= \int_{\left[t-y,t-y+\epsilon\right]}{(y+z)^{-1}f_{\rho}(z)\textrm{d}z}\nonumber\\
& \stackrel{\longrightarrow}{}\epsilon\frac{f_{\rho}(t-y)}{t}\hspace{4mm}\textrm{as} \hspace{4mm}\epsilon\rightarrow 0+\nonumber.
\end{align}
Again, we use the fact that
\begin{align}
\stackrel{\textrm{lim}}{_{\epsilon\rightarrow 0+}}\frac{1}{\epsilon}\int_{\left[x,x+\epsilon\right]}{p(z)\textrm{d}z}&= \int\delta_{z}(x)p(z)dz = p(x)\nonumber
\end{align}
to obtain the denominator and hence
\begin{align}
\mathbb{E}\left(W_1^*\in \textrm{d}\omega \mid T\in\left[t,t+\epsilon\right]\right) & \longrightarrow \frac{\omega \rho(\textrm{d}\omega)f_{\rho}(t-\omega)}{tf_{\rho}(t)}
\hspace{4mm}\textrm{as} \hspace{4mm}\epsilon\rightarrow 0+\nonumber.
\end{align}
\item \textbf{Suppose the statement holds true for $k>1$}, i.e.
\begin{align}
\mathbb{E}\left(W_k^*\in \textrm{d}\omega \mid T\in\left[t,t+\epsilon\right], W_j^*\in \textrm{d}\omega_j \forall j\in[k-1]\right) &=\nonumber\\
\mathbb{E}\left(W_k^*\in \textrm{d}\omega \mid T-\sum_{j=1}^{k-1}{W_j}\in\left[t-\sum^{k-1}{d\omega_j},t-\sum^{k-1}{d\omega_j}+\epsilon\right]\right)\nonumber \\
\frac{\omega f_{\rho}(t-\sum_{j=1}^{k}{\omega_j})}{(t-\sum_{j=1}^{k-1}{\omega_j})f_{\rho}(t-\sum_{j=1}^{k-1}{\omega_j})}\rho(\textrm{d}\omega)\nonumber
\hspace{4mm}\textrm{as} \hspace{4mm}\epsilon\rightarrow 0+.
\end{align}
\item \textbf{Case $k+1$}. 
\begin{align}
\mathbb{E}\left(W_{k+1}^*\in \textrm{d}\omega \mid T\in\left[t,t+\epsilon\right], W_j^*\in \textrm{d}\omega_j \forall j\in[k]\right) & = \nonumber\\
 \frac{\mathbb{E}\left(W_{k+1}^*\in \textrm{d}\omega, W_j^*\in \textrm{d}\omega_j ,\forall j\in[k],T\in\left[t,t+\epsilon\right]\right)}{\mathbb{E}\left(W_j^*\in \textrm{d}\omega_j, \forall j\in[k],T\in\left[t,t+\epsilon\right]\right)}\nonumber.
\end{align}
The denominator is given by the induction hypothesis (IH) so it's enough to prove this for the numerator:
\begin{align}
\mathbb{E}\left(W_{k+1}^*\in \textrm{d}\omega, W_j^*\in \textrm{d}\omega_j ,\forall j\in[k],T\in\left[t,t+\epsilon\right]\right)&=\nonumber \\
\mathbb{E}\left(\sum_{k=1}^{\infty}{\frac{\omega_k}{T}}\delta_{\omega_k}(\textrm{d}\omega_j),\forall j\in[k+1],T\in\left[t,t+\epsilon\right])\right)&\stackrel{\textrm{IH}}{=}\nonumber\\
  \int{\rho(y)y\delta_y(dw_{k+1})\mathbb{E}\left((y+T-\sum_{j=1}^{k}{W_j})^{-1},y+T-\sum_{j=1}^{k}{W_j} \in \left[t-\sum_{j=1}^{k}{\omega_j},t-\sum_{j=1}^{k}{\omega_j}+\epsilon\right]\right)}& \rightarrow\nonumber\\
 \frac{\epsilon}{(t-\sum_{j=1}^{k}{\omega_j})}\rho(\textrm{d}\omega)\omega f_{\rho}(t-\sum_{j=1}^{k+1}{\omega_j})\hspace{4mm}\textrm{as} \hspace{4mm}\epsilon\rightarrow 0+\nonumber
\end{align} 
Where, again, we use the IH in the second equality. Specifically, the fact the sequence of the first $k$ surpluses masses has the Markov property so it is enough to condition on the last surplus mass. Finally, we obtain:

\begin{align}
\mathbb{E}\left(W_{k+1}^*\in \textrm{d}\omega \mid T\in\left[t,t+\epsilon\right], W_j^*\in \textrm{d}\omega_j \forall j\in[k]\right) & =\nonumber\\
 \frac{\omega f_{\rho}(t-\sum_{j=1}^{k+1}{\omega_j})}{(t-\sum_{j=1}^{k}{\omega_j}) f_{\rho}(t-\sum_{j=1}^{k}{\omega_j})}\rho(\textrm{d}\omega)\hspace{4mm}\textrm{as} \hspace{4mm}\epsilon\rightarrow 0+\nonumber.
\end{align}
\end{enumerate}
\end{proof}

 \end{document}